\documentclass[a4paper,UKenglish,cleveref, autoref, thm-restate]{lipics-v2021}

\usepackage[utf8]{inputenc}
\usepackage{academicons}
\usepackage{amsmath}
\usepackage{amssymb}
\usepackage{doi}
\usepackage{thmtools}
\usepackage{enumerate}
\usepackage{graphicx}
\usepackage{lineno}
\usepackage{microtype}
\usepackage[numbers]{natbib}
\usepackage[parfill]{parskip}
\newcommand{\classt}[2]{\mathcal{C}_{\textnormal{#1}}(#2)}
\newcommand{\edget}[3]{\mathcal{E}_{\textnormal{#1}}(#2,#3)}
\newcommand{\classesx}[2]{\mathcal{S}_{\textnormal{#1}}(#2)}
\newcommand{\flipg}[2]{\mathcal{M}_{\textnormal{#1}}(#2)}
\newcommand{\tw}{\operatorname{tw}}
\newcommand{\cw}{\operatorname{cw}}
\newenvironment{proofsketch}{\renewcommand{\proofname}{Proof Sketch}\proof}{\endproof}

\hideLIPIcs  

\title{Rapid mixing for the hardcore Glauber dynamics and other Markov chains in bounded-treewidth graphs} 

\titlerunning{Rapid mixing for Glauber dynamics in bounded-treewidth graphs} 

\author{David Eppstein}{Department of Computer Science, University of California, Irvine, United States}{eppstein@uci.edu}{}{} 

\author{Daniel Frishberg}{Department of Computer Science and Software Engineering, California Polytechnic State University, San Luis Obispo, United States}{dfrishbe@calpoly.edu}{https://orcid.org/0000-0002-1861-5439}{}

\authorrunning{D. Eppstein and D. Frishberg} 

\Copyright{David Eppstein and Daniel Frishberg} 



\category{} 

\acknowledgements{The authors wish to acknowledge a number of helpful conversations on this topic with Milena Mihail, Ioannis Panageas, Eric Vigoda, Charlie Carlson, and Zongchen Chen, as well as with Karthik Gajulapalli, Hadi Khodabande, and Pedro Matias.}

\nolinenumbers 

\EventEditors{John Q. Open and Joan R. Access}
\EventNoEds{2}
\EventLongTitle{42nd Conference on Very Important Topics (CVIT 2016)}
\EventShortTitle{CVIT 2016}
\EventAcronym{CVIT}
\EventYear{2016}
\EventDate{December 24--27, 2016}
\EventLocation{Little Whinging, United Kingdom}
\EventLogo{}
\SeriesVolume{42}
\ArticleNo{23}

\begin{document}

\maketitle

\begin{abstract}
We give a new rapid mixing result for a natural random walk on the independent sets of a graph~$G$. We show that when $G$ has bounded treewidth, this random walk\textemdash known as the \emph{Glauber dynamics} for the \emph{hardcore model}\textemdash mixes rapidly for all fixed values of the standard parameter $\lambda > 0$, giving a simple alternative to existing sampling algorithms for these structures. We also show rapid mixing for analogous Markov chains on dominating sets, $b$-edge covers, $b$-matchings, maximal independent sets, and maximal $b$-matchings. (For $b$-matchings, maximal independent sets, and maximal $b$-matchings we also require bounded degree.) Our results imply simpler alternatives to known algorithms for the sampling and approximate counting problems in these graphs. We prove our results by applying a divide-and-conquer framework we developed in a previous paper, as an alternative to the \emph{projection-restriction} technique introduced by Jerrum, Son, Tetali, and Vigoda. We extend this prior framework to handle chains for which the application of that framework is not straightforward, strengthening existing results by Dyer, Goldberg, and Jerrum and by Heinrich for the Glauber dynamics on $q$-colorings of graphs of bounded treewidth and bounded degree.
\end{abstract}

\section{Introduction}
The \emph{Glauber dynamics} on independent sets in a graph\textemdash motivated in part by modeling systems in statistical physics\textemdash is a Markov chain in which one starts at an arbitrary independent set, then repeatedly chooses a vertex at random and, with probability that depends on a fixed parameter $\lambda > 0$, either removes the vertex from the set (if it is in the set), or adds it to the set (if it is not in the set and has no neighbor in the set). This chain, which samples from the \emph{hardcore model} on independent sets, has seen recent rapid mixing results under various conditions. In addition to independent sets, similar dynamics have been studied for a number of other structures\textemdash including, for example, $q$-colorings, matchings, and edge covers (more generally, $b$-matchings and $b$-edge covers). 

\subsection{Our contribution}
We prove that the hardcore Glauber dynamics mixes rapidly on graphs of bounded treewidth for all fixed $\lambda > 0$, and that the Glauber dynamics on partial $q$-colorings (for all $\lambda > 0$) of a graph of bounded treewidth, and on $q$-colorings of a graph of bounded treewidth and degree, mix rapidly. Marc Heinrich proved the latter result, namely for $q$-colorings, in a 2020 preprint~\cite{heinrich2020glauber}. Heinrich's result applies to all graphs of bounded treewidth; however, for graphs of bounded treewidth and degree, whose degree is less than quadratic in their treewidth, we improve on Heinrich's upper bound\textemdash provided that $q$ is fixed. We also prove that the analogous dynamics on the $b$-edge covers (when $b$ is bounded) and the dominating sets of a graph of bounded treewidth mix rapidly for all $\lambda > 0$. In a similar vein, we prove that three additional chains\textemdash on $b$-matchings (when $\lambda > 0$), on maximal independent sets, and on maximal $b$-matchings\textemdash mix rapidly in graphs of bounded treewidth and degree.

To prove our results, we apply a framework we introduced in a companion paper~\cite{eppfrishtri} that uses the \emph{multicommodity flow} technique (essentially the same as the \emph{canonical paths} technique) for bounding mixing times. (We previously presented this framework in a preprint of the present paper~\cite{eppfrisharx}.) The framework consists of a set of conditions (which we will define in Section~\ref{sec:fwcond}) that guarantee rapid mixing; these conditions make progress towards unifying prior work on similar Glauber dynamics with prior work on probabilistic graphical models. In that paper~\cite{eppfrishtri}, we also proved that the \emph{flip walk} on the \emph{$k$-angulations} of a convex $n$-point set mixes in time quasipolynomial in $n$ for all fixed $k \geq 3$, although the special case $k=3$ was known already to mix rapidly~\cite{mct}. Thus our framework applies beyond graphical models and graph sampling problems.

\subsection{Main results}
Our main results are the following (see Section~\ref{sec:prelim} for relevant definitions). 

\begin{theorem}
\label{thm:indsetmix}
The hardcore Glauber dynamics mixes in time
$n^{O(t)}$
on graphs of treewidth~$t$ for all fixed $\lambda > 0$.
\end{theorem}

\begin{theorem}
\label{thm:lcolor}
The (unbiased) Glauber dynamics on $q$-colorings (when $q \geq \Delta + 2$ is fixed) mixes in time
$n^{O(t)}$ on graphs of treewidth~$t$ and bounded degree when~$q$ is fixed. The Glauber dynamics on partial $q$-colorings (when $q \geq \Delta + 2$ is fixed) mixes in time
$n^{O(t)}$ on graphs of treewidth~$t$ for all fixed~$\lambda > 0$.
\end{theorem}

\begin{theorem}
\label{thm:domsetmix}
The Glauber dynamics on $b$-edge covers mixes in time
$n^{O(t^2)}$
on graphs of treewidth~$t$, for all fixed~$b \geq 1$ and fixed~$\lambda > 0$. The Glauber dynamics on dominating sets mixes in time
$n^{O(t)}$
on graphs of treewidth~$t$ for all fixed~$\lambda > 0$. The Glauber dynamics on $b$-matchings mixes in time 
$n^{O(t)}$
on graphs of treewidth~$t$ and bounded degree~$\Delta$ for all fixed~$\lambda > 0$ and fixed~$b \geq 1$.
\end{theorem}

\begin{theorem}
\label{thm:mindsetmix}
There exist Markov chains on maximal independent sets and maximal $b$-matchings, whose stationary distributions are uniform, that mix in time
$n^{O(t)}$
on graphs of treewidth~$t$ and bounded degree.
\end{theorem}

\subsection{The framework: recursive flow construction}
A \emph{multicommodity flow} in an undirected graph $G = (V, E)$ with~$n$ vertices is a set of $n^2$ flows, one flow for each ordered pair of vertices $(s, t)$, where each flow sends one unit of a commodity from $s$ to $t$. More precisely, take each (undirected) edge in~$E$ and make two directed copies, one in each direction; let~$E^+ = \bigcup_{\{u,v\}\in E} \{(u, v), (v, u)\}$ denote the set of all these directed copies. A multicommodity flow is a collection of functions $f_{st}: E^+ \rightarrow \mathbb{R}_{\geq 0}$ such that each~$f_{st}$ is a valid flow function, with~$s$ (respectively~$t$) having net out flow (respectively in flow) equal to one, and all other vertices having zero net flow. If a multicommodity flow exists in $G$ with small \emph{congestion}\textemdash i.e. one in which no edge carries too much flow\textemdash then the natural Markov chain whose states are the vertices of $G$ mixes rapidly.

The chains we analyze are natural random walks on a \emph{Glauber graph}~$\flipg{}{G}$\footnote{The chains on maximal independent sets and maximal $b$-matchings are not strictly Glauber dynamics, but we will use the same term for the graph, redefining the edge set as pairs connected by the moves we define in Appendix~\ref{sec:dombmax}.}\textemdash the graph whose vertices are the structures over which the random walk is performed, and whose edges are the pairs of these structures with symmetric distance equal to one. For example, in the case of independent sets, $\flipg{}{G}$ has as its vertex set the collection of all independent sets in~$G$, and as its edge set the collection of all (unordered) pairs of independent sets~$S, S'$ in~$G$ such that $S = S' \cup \{v\}$ for some~$v\in V(G)$. Thus each of these random walks is performed on a graph that may be exponentially large with respect to the size of the input graph. In our previous work~\cite{eppfrishtri}, we showed that when all of a certain set of conditions hold, we can construct a multicommodity flow in $\flipg{}{G}$ with congestion polynomial in $n = |V(G)|$, implying that the walk on $\flipg{}{G}$ mixes rapidly. The conditions specify that $\flipg{}{G}$ can be partitioned into a small number of induced subgraphs, all of which are approximately the same size, with large numbers of edges between pairs of the subgraphs. The conditions require that each of these induced subgraphs can be decomposed into smaller Glauber graphs that are similar in structure to $\flipg{}{G}$. This self similarity allows for the recursive construction of a multicommodity flow, by assembling flows on smaller Glauber graphs together into a flow in $\flipg{}{G}$ with small congestion.

\subsection{Projection-restriction and prior work on the hardcore model}
\label{sec:priorsim}
Prior work on rapid mixing of Markov chains on subset systems includes the special case of matroid polytopes. For this case, recent results~\cite{anari1,anari2} have partly solved a 30-year-old conjecture of Mihail and Vazirani~\cite{mihail1989expansion}. Other prior work uses multicommodity flows (and the essentially equivalent \emph{canonical paths} technique) to obtain polynomial mixing upper bounds on structures of exponential size, including matchings and 0/1 knapsack solutions~\cite{morris2004random, rapidsurvey}. Madras and Randall \cite{mrdecomp} used a decomposition of the hardcore model state space to prove rapid mixing under different conditions. We also decompose the state space, but our approach is different and is more similar to Heinrich's~\cite{heinrich2020glauber} application of the \emph{projection-restriction} technique introduced by Jerrum, Son, Tetali, and Vigoda~\cite{jerrumprojres}. This technique involves partitioning the state space of a chain into a collection of sub-state spaces, each of which internally has a good \emph{spectral gap}\textemdash a property that implies rapid mixing\textemdash and all of which are well connected to one another. Heinrich used the vertex separation properties of bounded-treewidth graphs to obtain an inductive argument: the resulting sub-spaces are themselves \emph{Cartesian products} of chains on smaller graphs, and thus mix rapidly. (See Lemma~\ref{lem:cartflow}.) We partition the state space recursively using the same vertex separation properties, and indeed for the chains on $b$-matchings and $q$-colorings in bounded-treewidth, bounded-degree graphs, combining these properties straightforwardly with the existing spectral projection-restriction machinery of~\cite{jerrumprojres} suffices for rapid mixing. The main contribution in this paper is to extend the framework to chains for which this application is not straightforward. That is, we give general conditions for constructing a multicommodity flow in the projection chain with small congestion, giving a good spectral gap in the projection chain. One may then apply induction using the spectral machinery of~\cite{jerrumprojres} to obtain rapid mixing in the overall chain; alternatively, one can substitute the flow-based machinery from our companion paper~\cite{eppfrishtri} for the spectral technique.

In the case of independent sets, Jerrum, Son, Tetali, and Vigoda~\cite{jerrumprojres} applied their technique to a special case of the hardcore model, namely regular trees. However, it was not clear how to generalize this application to bounded-treewidth graphs\textemdash since showing the spectral gap of the projection chain is sufficiently large is not straightforward. Martinelli, Sinclair, and Weitz~\cite{martinellitrees} showed that the Glauber dynamics on the hardcore model mixes in $O(n\log n)$ time on the complete $\Delta - 1$-ary tree with $n$ nodes, but they did not address general trees. Berger, Kenyon, Mossel, and Peres~\cite{bkmp} showed rapid mixing for $q$-colorings of regular trees with unbounded degree but also did not address general trees. Our first main technical contribution is to show rapid mixing for general bounded-treewidth graphs by introducing the \emph{hierarchical} version of our framework, in which we construct a flow with small congestion in the projection chain; we show that this construction gives rapid mixing for dominating sets, \emph{partial} $q$-colorings, and $b$-edge covers in bounded-treewidth graphs. We solve another problem: the technical theorem in~\cite{jerrumprojres} as stated requires each of the state spaces in the partition to be a Cartesian product of chains on smaller spaces. For four of our eight chains\textemdash those on dominating sets, $b$-edge covers, maximal $b$-matchings, and maximal independent sets\textemdash the sub-spaces obtained in the decomposition are not a disjoint union of Cartesian products but may each be a union of Cartesian products, or may be mutually intersecting. In some cases, the sub-spaces may even induce disconnected restriction chains. Our second main contribution is to resolve this problem, using the structure of the state spaces of Glauber dynamics as graphs. We discuss this in Appendix~\ref{sec:conclusion}.

\subsection{Paper organization}
In Section~\ref{sec:prelim}, we give relevant background. In Section~\ref{sec:fwcond}, we use the chain on independent sets to review the ``non-hierarchical'' version of our framework (the version we gave in our companion paper)\textemdash which works for this chain when treewidth and degree are bounded. In Sections~\ref{sec:nonhierqcol} and~\ref{sec:bprobs} we apply it to $q$-colorings and to $b$-edge covers and $b$-matchings. To fully prove Theorem~\ref{thm:indsetmix} and Theorem~\ref{thm:domsetmix}, we need to deal with unbounded-degree graphs\textemdash our first main technical contribution. In Section~\ref{sec:altfwintro}, we modify the framework to do so, proving Theorem~\ref{thm:indsetmix} for $\lambda = 1$. We defer some details to Appendix~\ref{sec:mainfwcond}, where we also finish the proof of Theorem~\ref{thm:lcolor} for $\lambda = 1$. We prove the general case $\lambda > 0$ of Theorems~\ref{thm:indsetmix} and \ref{thm:lcolor} in Appendix~\ref{sec:alllambda}. We finish the proofs of Theorems~\ref{thm:domsetmix} and~\ref{thm:mindsetmix} in Appendix~\ref{sec:adapting}: applying the framework to the relevant chains requires a further refinement of the framework. In all of the above, we prove rapid mixing but defer derivation of specific upper bounds to Section~\ref{sec:mixingder}.

\section{Preliminaries}
\label{sec:prelim}
\subsection{Glauber dynamics}
\label{sec:mixingexp}
\begin{definition}
\label{def:hardcore}
The \emph{hardcore Glauber dynamics} on a graph $G$ is the following chain, defined with respect to a fixed real parameter $\lambda > 0$:
\begin{enumerate}
\item Let $X_0$ be an arbitrary independent set in $G$.
\item For $t \geq 0$, select a vertex $v \in V(G)$ uniformly at random.
\item If $v \notin X_t$ and $X_t \cup \{v\}$ is not a valid independent set, do nothing.
\item Otherwise:
\subitem Let $X_{t+1} = X_t \cup \{v\}$ with probability $\lambda/(\lambda + 1)$.
\subitem Let $X_{t+1} = X_t \setminus \{v\}$ with probability $1/(\lambda + 1)$.
\end{enumerate}
\end{definition}

Graph-theoretically, the Glauber dynamics is defined as follows: let the \emph{indepdendent set Glauber graph}~$\flipg{IS}{G}$ denote the graph whose vertices are identified with the independent sets of a given graph~$G$, and whose edges are the pairs of independent sets whose symmetric difference is one. The hardcore Glauber dynamics is a Markov chain, parameterized by $\lambda > 0$, with state space $\Omega = V(\flipg{IS}{G})$ and probability matrix $P$, where for $S, S' \in V(\flipg{}{G})$ with $S \neq S'$, $P(S, S') = \lambda/(\Delta_{\mathcal{M}}(\lambda + 1))$
when $|S' \setminus S| = 1$, and
$P(S, S') = 1/(\Delta_{\mathcal{M}}(\lambda + 1))$
when $|S \setminus S'| = 1$. If $S = S'$, then $P(S, S') = 1 - \sum_{S''\neq S}P(S, S'')$.
(Here $\Delta_{\mathcal{M}}$ is the maximum degree of the Glauber graph\textemdash i.e. the maximum number of neighboring states that a state $S$ can have.)
The Glauber graph has vertex set~$\Omega$ and adjacency matrix~$P$\textemdash up to the addition of self loops and normalization by degree. (When $\lambda \neq 1$ this graph can still be augmented with suitable weights so that the walk on the graph is the Glauber dynamics.)

\subsection{Mixing time}
\label{sec:prelimmixingexp}
To generate, approximately uniformly at random, an object of a given class\textemdash such as an independent set in a given graph\textemdash one can conduct a random walk on a graph whose vertices are the objects of interest, and whose edges represent local moves between the objects (or states). It is known that under certain mild conditions satisfied by as all our chains (see Appendix~\ref{sec:prelimrev}), the walk converges to the uniform distribution in the limit. The rate of convergence is important: in the case of subset systems such as those we consider, the walk takes place over an exponentially large number of subsets defined over an underlying set of size $n$. If the convergence, or \emph{mixing time}, of the walk is polynomial in $n$, then the random walk is said to be \emph{rapidly mixing}. The mixing time is denoted~$\tau(\varepsilon)$, where $\varepsilon$ denotes the desired precision of convergence to the uniform distribution, and the value of $\tau$ at $\varepsilon$ is the minimum number of steps in the random walk before convergence is guaranteed. Convergence is measured via the \emph{total variation distance} \cite{sinclair_1992} between the distribution over states induced by the walk at a given time step, and the uniform distribution. One can obtain non-uniform \emph{stationary} distributions by weighting the graph\textemdash see Appendix~\ref{sec:prelimrev}. See Levin, Peres, and Wilmer~\cite{levin2009markov} for a comprehensive treatment of rapid mixing.

A Markov chain, given a starting state $S \in \Omega$, induces a probability distribution $\pi_t$ at each time step $t$. The Glauber dynamics is known, regardless of starting state, to converge in the limit to a \emph{stationary} distribution $\pi^*(S) = \lambda^{|S|}/Z(\flipg{}{G}),$ where the term $Z(\flipg{}{G})$ is simply a normalizing value. When $\lambda$ is unspecified, assume $\lambda = 1$ (the uniform case). The \emph{mixing time} is defined as follows:

Given an arbitrary $\varepsilon > 0$, the \emph{mixing time}, $\tau(\varepsilon)$, of a Markov chain with state space $\Omega$ and stationary distribution $\pi^*$ is the minimum time $t$ such that, regardless of starting state, we always have
$\frac{1}{2}\sum_{S \in \Omega}|\pi(S) - \pi^*(S)| < \varepsilon.$
Suppose the chain belongs to a family of Markov chains, the size of whose state space is parameterized by some value $n$. Here, $|\Omega|$ may be exponential in $n$. If $\tau(\varepsilon)$ is bounded by a polynomial function in $\log (1/\varepsilon)$ and in $n$, the chain is said to be \emph{rapidly mixing}. It is common to omit the parameter $\varepsilon$ and assume~$\varepsilon = 1/4$.

\subsection{Treewidth and vertex separators}
\begin{definition}\cite{robertsontw}
\label{def:treedecomp}
A \emph{tree decomposition} of a graph $G = (V, E)$ is a collection of sets $\{X_i\}, i = 1, \dots, k$, called \emph{bags}, together with a tree $T$, whose nodes are identified with the bags $\{X_i\}$, such that all of the following hold:
\begin{enumerate}
\item Every vertex in $V$ lies in some bag, i.e. $\bigcup_{i=1}^k X_i = V$.
\item For every $(u, v) \in E$, the vertices $u$ and $v$ belong to at least one bag $X_i$ together, i.e. for some $i$, $u \in X_i$ and $v \in X_i$.
\item The collection of all bags containing any given vertex $v \in V$, i.e. $\{X_i \mid v \in X_i\}$ forms a (connected) subtree of $T$.
\end{enumerate}
\end{definition}

\begin{definition}\cite{robertsontw}
\label{def:treewidth}
The \emph{width} of a tree decomposition is one less than the size of the largest bag in the decomposition. The \emph{treewidth} of a graph $G$ is the minimum $t$ such that a tree decomposition of $G$ exists with width $t$.
\end{definition}

Intuitively, treewidth measures how far away a graph is from being a tree. For example, trees have treewidth one; a graph consisting of a single cycle of size at least three has treewidth two. Treewidth is of interest in large part because many NP-hard problems become tractable on graphs of bounded treewidth. For a full definition of treewidth and a survey of this phenomenon, known as \emph{fixed-parameter tractability}, see \cite{FPTsurvey}.

For our purposes, treewidth is of interest due to its relationship to \emph{vertex separators}: a vertex set~$X \subseteq V$ in a graph~$G = (V, E)$ is called a \emph{vertex separator} if the deletion of~$X$ from~$V$ leaves the induced subgraph on the remaining vertices disconnected. Say that~$X$ is a \emph{balanced separator} if deleting~$X$ partitions~$V$ into mutually disconnected subsets~$A \cup B = V \setminus X$ such that~$|V|/3 \leq |A| \leq |B| \leq 2|V|/3$. A graph~$G$ is \emph{recursively $s$-separable}~\cite{ericksontw} if either (i) $|V(G)| \leq 1$, or (ii) $G$ has a balanced separator $X$ with~$|X| \leq s$ and, after deleting~$X$, the resulting subsets~$A$ and~$B$ induce subgraphs of~$G$ that are each recursively $s$-separable.

The following is known and easy to prove \cite{ericksontw}:
\begin{lemma}
\label{lem:vseptw}
Every graph with treewidth $t \geq 1$ is recursively $s$-separable for all~$s \geq t + 1$.
\end{lemma}

\section{$\lambda = 1$: Bounded treewidth and degree}
\label{sec:fwcond}
To build up to the proof of Theorem~\ref{thm:indsetmix}, we first show a weaker result: that the unifrom hardcore Glauber dynamics mixes rapidly in graphs of bounded treewidth and degree. Fully proving Theorem~\ref{thm:indsetmix}, even in the unbiased case, requires the non-hierarchical framework. The main technical lemma in this section, Lemma~\ref{lem:condexp}, comes from our companion paper. Our contribution in this paper is the application to independent sets in graphs of bounded treewidth and degree\textemdash which we strengthen to graphs of bounded treewidth in Section~\ref{sec:altfwintro}.

The following is necessary for the Glauber dynamics to sample correctly:
\begin{lemma}
\label{lem:isflipconn}
The independent set Glauber graph is connected.
\end{lemma}
\begin{proof}
Consider the empty independent set $\emptyset$. Every independent set $S \in V(\flipg{IS}{G})$ has a path of length $|S|$ to $\emptyset$, formed by removing each vertex in $S$ in arbitrary order.
\end{proof}

\subsection{Partitioning the vertices of $\flipg{IS}{G}$ into classes}
\label{sec:fwbrief}

The vertices of the Glauber graph $\flipg{IS}{G}$ are subsets of the vertices of an underlying graph $G$. When $G$ has bounded treewidth, we can choose a small separator $X$ that partitions $V(G) \setminus X$ into two mutually disconnected vertex subsets, $A$ and $B$, each of which has at most~$2|V(G)|/3$ vertices. Consider the problem of sampling an independent set $S$ from $G$. Given a separator $X$ for $G$, partition the independent sets in $G$ into equivalence classes as follows:

\begin{definition}
\label{def:ispart}
Let $G = (V, E)$ be a graph. Let $\flipg{IS}{G}$ be the independent set Glauber graph we have defined. Let $X \subseteq V$ be a vertex separator for $G$. Let $\classesx{IS}{G}$ be the set of equivalence classes of $V(\flipg{IS}{G})$ in which two independent sets $S$ and $S'$ are in the same class if $S \cap X = S' \cap X$. Let $T = S \cap X$, and call the corresponding class $\classt{IS}{T}$.
\end{definition}
(Technically~$X$ is also a parameter for~$\classesx{IS}{G}$ and~$\classt{IS}{T}$, but we omit it for ease of notation.)

\begin{figure}
\centering
\includegraphics[width=15em]{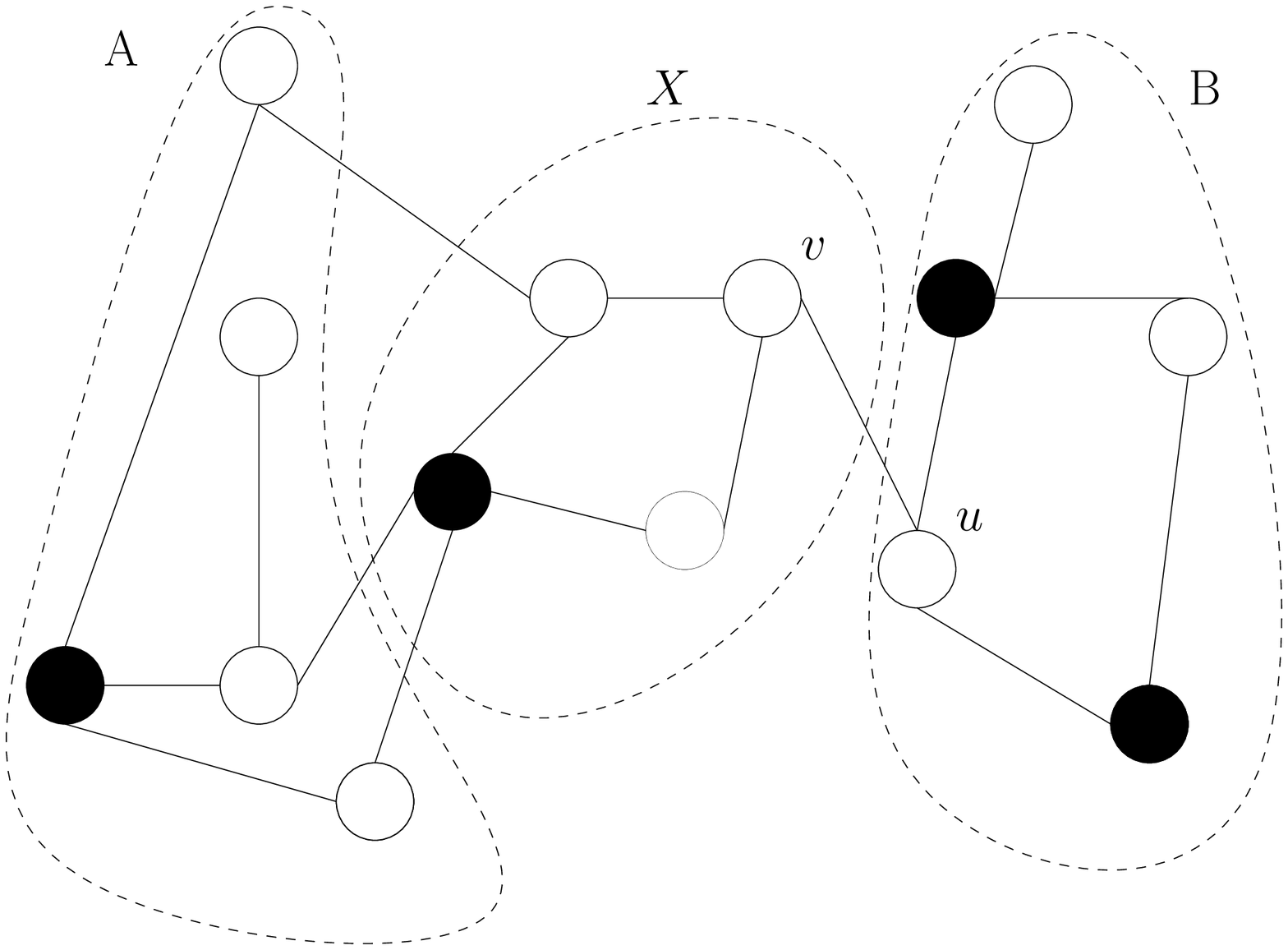}\hspace{1cm}
\includegraphics[width=15em]{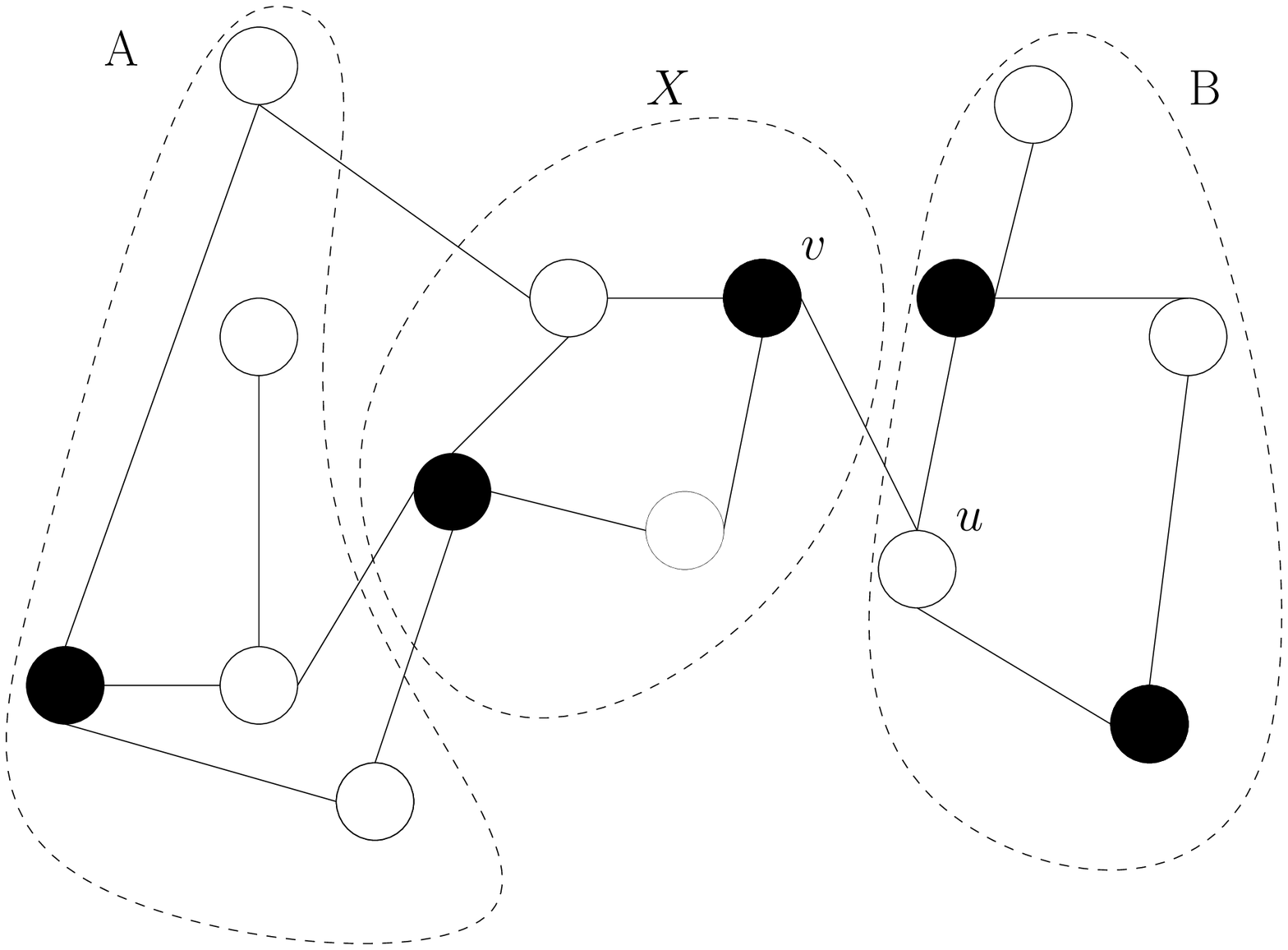}
\caption{Two independent sets in a graph $G$: $S$ (left) and $S'$ (right), belonging to distinct classes. $S$ and $S'$ differ by a move, with the separator~$X$ inducing the classes to which the sets belong. $S'$ results from adding $v$ to $S$. $|S'| < |S|$, since $S'$ excludes those independent sets that contain~$u$.}
\label{fig:isflip}
\end{figure}

See Figure~\ref{fig:isflip} for an example of a partitioning.

The \emph{Cartesian product} of two graphs~$H$ and~$J$ is the graph whose vertex set is~$V(H) \times V(J)$ and whose edges are the pairs $((h_1, j_1), (h_2, j_2))$ such that either $h_1 = h_2$ and $(j_1, j_2) \in E(J)$ or else $(h_1, h_2) \in E(H)$ and $j_1 = j_2$.

Let~$A$ and~$B$ be the mutually disconnected vertex subsets into which the removal of~$X$ partitions $V(G) \setminus X$. Given a fixed independent subset~$T \subseteq X$, identify the independent sets in~$\classt{IS}{T}$ with the pairs of the form~$(S_A, S_B)$, where~$S_A$ is an independent set in~$A \setminus N_A(T)$, and~$S_B$ is an independent set in~$B \setminus N_B(T)$, where~$N_A(T)$ and~$N_B(T)$ denote the union of the neighborhoods of vertices in~$T$, in~$A$ and~$B$ respectively. That is, identify each independent set in~$\classt{IS}{T}$ with a pair of an independent set in~$A$ that avoids neighbors of vertices in~$T$, and a similar independent set in~$B$. Consider the two Glauber graphs~$\flipg{IS}{A\setminus N_A(T)}$ and~$\flipg{IS}{B\setminus N_B(T)}$, whose vertices are respectively the independent sets in~$G[A \setminus N_A(T)]$, and those in~$G[B \setminus N_B(T)]$. If two independent sets $S = (S_A, S_B)$ and $S' = (S_A', S_B')$ belong to the same class, then a \emph{move} (traversal of an edge in the Glauber graph) exists between~$S$ and~$S'$ in~$\flipg{IS}{G}$ precisely when a move exists between the restrictions of $S$ and $S'$ to either~$\flipg{IS}{A\setminus N_A(T)}$ or~$\flipg{IS}{B\setminus N_B(T)}$ (but not both). Therefore, each class induces, in~$\flipg{IS}{G}$, a subgraph that is isomorphic to a Cartesian product of two smaller Glauber graphs: 

\begin{lemma}
\label{lem:classtcart}
Given a graph~$G$ and a vertex separator~$X$ that partitions~$V(G)$ into subgraphs~$A$ and~$B$, for every class~$T \in \classesx{IS}{G}$, $\classt{IS}{T} \cong \flipg{IS}{A\setminus N_A(T)} \Box \flipg{IS}{B\setminus N_B(T)}.$
\end{lemma}
(Here by the symbol~$\cong$ we denote isomorphism, and we identify the class~$\classt{IS}{T}$ with the subgraph it induces in~$\flipg{IS}{G}$.)

\subsection{Rapid mixing for the hardcore Glauber dynamics when $G$ has bounded treewidth and degree}
\begin{figure}
\centering
\includegraphics[width=25em]{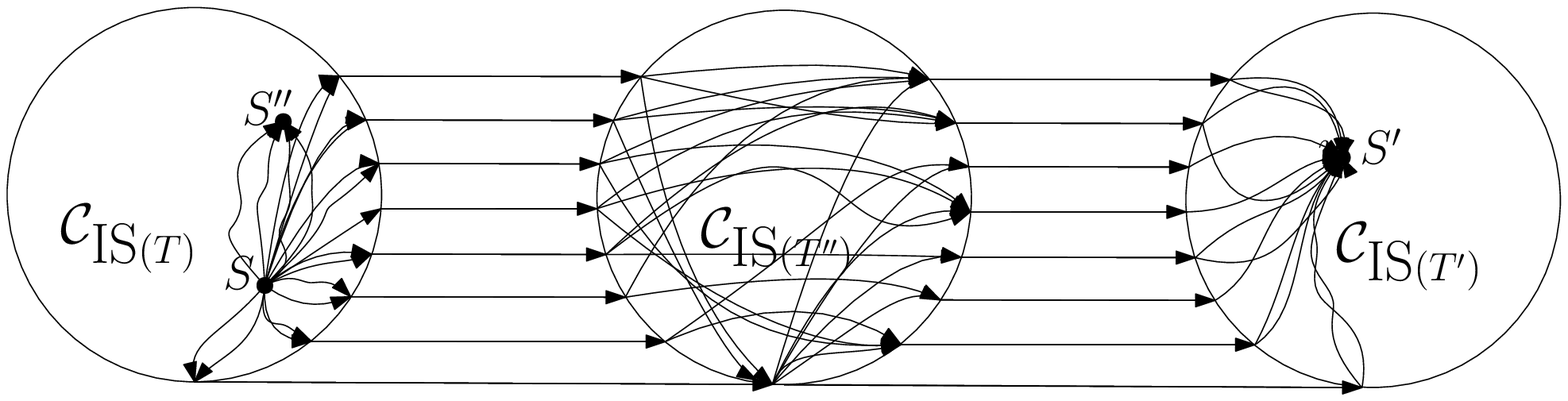}
\caption{A schematic view of three classes in the independent set Glauber graph $\flipg{IS}{G}$. The large circles denote classes under the partition described in Section~\ref{sec:fwbrief}. The curved arrows illustrate the construction of a flow in $\flipg{IS}{G}$ from an independent set $S \in \classt{IS}{T}$ to another independent set $S'' \in \classt{IS}{T}$\textemdash and also to an independent set $S' \in \classt{IS}{T'}$. Here, $\classt{IS}{T}$ and $\classt{IS}{T''}$ are adjacent classes in $\flipg{IS}{G}$, connected by a large number of edges, and similarly $\classt{IS}{T'}$ and $\classt{IS}{T''}$ are adjacent. In Section~\ref{sec:iscond} we formalize this flow.}
\label{fig:mgraphflow}
\end{figure}
As described in Section~\ref{sec:fwbrief}, we use a small vertex separator $X$ in $G$ to give a decomposition of $\flipg{IS}{G}$ into subgraphs, each of which has a Cartesian product structure\textemdash in which both factor graphs in the product are themselves Glauber graphs. Since Cartesian products preserve flow congestion upper bounds (see Lemma~\ref{lem:cartflow}), this decomposition provides a crucial inductive structure. We analyze this structure in this section.

\label{sec:iscond}

\begin{restatable}{lemma}{lemiscondfacts}
\label{lem:iscondfacts}
\label{lem:iscondnumclasses}
\label{lem:iscondclasssizes}
\label{lem:iscondsedeg}
\label{lem:iscondsesizes}
Let $G$ be a graph with bounded treewidth $t$ and bounded degree $\Delta$, let $\flipg{IS}{G}$ be as we have defined, and let $\classesx{IS}{G}$ be as in Definition~\ref{def:ispart} with respect to a small balanced separator~$X$ with $|X| \leq t + 1$. Then:
\begin{enumerate}
\item The number of classes in $\classesx{IS}{G}$ is $O(1)$.
\item For every pair of classes $\classt{IS}{T}, \classt{IS}{T'} \in \classesx{IS}{G}$, $|\classt{IS}{T}| = \Theta(1)|\classt{IS}{T'}|$.
\item Let $\classt{IS}{T}, \classt{IS}{T'} \in \classesx{IS}{G}$ be two classes. No independent set $S \in \classt{IS}{T}$ has more than one move to an independent set $S' \in \classt{IS}{T'}$.
\item Let $\classt{IS}{T}, \classt{IS}{T'} \in \classesx{IS}{G}$ be two classes. Suppose there exists at least one move between an independent set in $\classt{IS}{T}$ and an independent set in $\classt{IS}{T'}$. Then there exist at least $\Omega(1)|\classt{IS}{T}|$ moves between independent sets in $\classt{IS}{T}$ and independent sets in $\classt{IS}{T'}$.
\end{enumerate}
\end{restatable}

\begin{proof}
Claim 1 follows from the fact that $|\classesx{IS}{G}| \leq 2^{|X|} \leq 2^{t+1} = O(1),$ where the first inequality is true because each class is identified with a subset of the vertices in $X$. 
The proofs of claims 2 through 4 are in Appendix~\ref{sec:appdetails}.
\end{proof}

We will use Lemma~\ref{lem:iscondfacts} to prove the following, applying the framework from our previous paper, in Section~\ref{sec:fwcond}:
\begin{restatable}{lemma}{lemiscondexp}
\label{lem:iscondexp}
Given a graph $G$ with bounded treewidth and degree, the natural random walk on the independent set Glauber graph $\flipg{IS}{G}$ has mixing time $\tau(\varepsilon) = O(n^c\log{1/\varepsilon})$, where $c = O(1)$.
\end{restatable}

To prove Theorem~\ref{thm:indsetmix}, however, we need to get rid of the assumption that degree is bounded. We address this issue in Section~\ref{sec:altfwintro}. 

\subsection{Abstraction into framework conditions}
\label{sec:fwcond}
The observations in Lemma~\ref{lem:iscondnumclasses} correspond to a set of conditions we gave in our previous work~\cite{eppfrishtri}. These conditions are, given a connected graph $\flipg{}{G}$, on some set of combinatorial structures over an underlying graph $G$ with $n$ vertices:
\begin{enumerate}
\item \label{cond1} The vertices of $\flipg{}{G}$ can be partitioned into a set $\mathcal{S}$ of classes, where $|\mathcal{S}| = O(1).$
\item \label{cond2} The ratio of the sizes of any two classes in $\mathcal{S}$ is $\Theta(1)$.
\item \label{cond3} Given two classes $\classt{}{T}, \classt{}{T'} \in \mathcal{S}$, no vertex in $\classt{}{T}$ has more than $O(1)$ edges to vertices in $\classt{}{T'}$.
\item \label{cond4} For every pair of classes that share at least one edge, the number of edges between the two classes is $\Theta(1)$ times the size of each of the two classes.
\item \label{cond5} Each class in $\mathcal{S}$ is the Cartesian product of two graphs $\flipg{}{G_1}$ and $\flipg{}{G_2}$, each of which can be recursively partitioned in the same way as $\flipg{}{G}$.
\item \label{cond6} The recursive partitioning mentioned in Condition~\ref{cond5} reaches the base case (graphs with one or zero vertices) in $O(\log n)$ steps.
\end{enumerate}

Conditions~\ref{cond1} through~\ref{cond4} correspond respectively to Lemma~\ref{lem:iscondnumclasses}; Condition~\ref{cond5} corresponds to Lemma~\ref{lem:classtcart}. Condition~\ref{cond6} corresponds to the observation at the end of the proof sketch of Lemma~\ref{lem:iscondexp}. 

We introduce some facts that we previously used to prove that these conditions suffice for rapid mixing, via \emph{expansion}, then review a sketch of the proof; we will build on these techniques in Section~\ref{sec:altfwintro} (our main contribution) and in the appendices.

The \emph{edge expansion} (or simply the \emph{expansion}) of a graph~$G = (V, E)$ is the quantity
$h(G) := \min_{S \subseteq V: |S| \leq |V|/2} \frac{|\{(u, v)\in E \mid u \in S, v \notin S\}|}{|S|},$
i.e. the minimum quotient of the number of edges in the cut by the number of vertices on the smaller side of the cut.
The \emph{vertex expansion} is the quantity
$h(G) := \min_{S \subseteq V: |S| \leq |V|/2} \frac{|\{v \in V \setminus S \mid \exists u \in S, (u, v) \in E\}|}{|S|},$
i.e. the minimum quotient of the number of neighbors of a set~$S$ with~$|S| \leq |V|/2$ that are not in~$S$.

Mixing can be bounded from above via a lower bound on expansion~\cite{sinclair_1992} when the degree of a Glauber graph is small (linear in the case of our chains):
\begin{lemma}
\label{lem:expmixing}
Given a graph~$\mathcal{M} = (\mathcal{V}, \mathcal{E})$, consider the Markov chain whose state space is~$\mathcal{V}$ and whose transitions are of the form~$P(x, x) = 1/2, P(x, y) = 0 \forall (x,y)\notin \mathcal{E}, P(x, y) = 1/(2\Delta) \forall (x, y) \in \mathcal{E}$, where~$\Delta$ is the maximum degree of~$\mathcal{M}$.
The mixing time of this Markov chain is at most
$$\tau(\varepsilon) = O\left(\frac{\Delta^2}{(h(\mathcal{M}))^2}\cdot \ln{\frac{|\mathcal{V}|}{\varepsilon}}\right).$$
\end{lemma}

Expansion, in turn, can be bounded from below via an upper bound on the \emph{congestion} of a multicommodity flow. Given a multicommodity flow~$f = \{f_{st} | s,t\in V\times V\}$ in a graph~$G = (V, E)$, define the congestion of~$f$ as the quantity
$\rho = \max_{(u,v)\in E^+} \sum_{s,t\in V\times V} f_{st}(u, v),$
i.e. the maximum amount of flow sent across an edge.
\begin{lemma}~\cite{sinclair_1992}:
\label{lem:flowexp}
For every graph $G = (V, E)$ and for every flow function~$f$ defined over~$G$ and having congestion~$\rho$, $h(G) \geq 1/(2\rho)$.
\end{lemma}
\begin{restatable}{lemma}{lemcartflow}
\label{lem:cartflow}
Given graphs $H$ and $J$, let~$G$ be the Cartesian product~$H \Box J$. Suppose multicommodity flows exist in~$H$ and~$J$ with congestion at most~$\rho_H$ and~$\rho_J$ respectively. Then there exists a multicommodity flow in~$G$ with congestion at most~$\max\{\rho_H, \rho_J\}$.
\end{restatable}
We proved Lemma~\ref{lem:cartflow} in~\cite{eppfrishtri}, although an analogous result for expansion is known~\cite{Graham1998IsoperimetricIF}. We also proved the following in~\cite{eppfrishtri}. We review the lemma and give a modified proof sketch here, in terms more intuitive for the chains we are analyzing in this paper. We will modify the technique in Section~\ref{sec:altfwintro}.
\begin{restatable}{lemma}{lemcondexp}
\label{lem:condexp}
Given a graph $\flipg{}{G}$ satisfying the conditions in Section~\ref{sec:fwcond}, the expansion of $\flipg{}{G}$ is $\Omega(1/n^c)$, where $c = O(1).$ 
\end{restatable}
\begin{proofsketch}
Partition~$\flipg{}{G}$ into classes as in Definition~\ref{def:ispart}. By Lemma~\ref{lem:classtcart}, each class $\classt{}{T} \in \classesx{}{G}$ is isomorphic to the Cartesian product $\flipg{}{A\setminus N_A(T)} \Box \flipg{}{B\setminus N_B(T)}$. We make an inductive argument, in which the inductive hypothesis assumes that for each such Cartesian product, the graphs $\flipg{}{A\setminus N_A(T)}$ and $\flipg{}{B \setminus N_B(T)}$ have multicommodity flows $f_A$ and $f_B$ with congestion~$\rho_{A} \leq c^{\log |V(G)|-1}, \rho_{B} \leq c^{\log |V(G)|-1}$
respectively, for some constant $c$. By Lemma~\ref{lem:cartflow}, $\classt{}{T}$ then has a flow $f_T$ with congestion~$\rho_{T} \leq c^{\log |V(G)|-1}.$ The inductive step is to combine the $f_T$ flows for all of the classes, giving a flow $f$ in $\flipg{}{G}$ with small congestion. We need to route flow between every $S, S' \in V(\flipg{}{G})$. If $S$ and $S'$ belong to the same class $\classt{}{T}$, simply use the same flow that $S$ uses to send its unit to $S'$ in $f_T$. If $S\in \classt{}{T}$ and $S'\in \classt{}{T'} \neq \classt{}{T}$ belong to different classes, we find a sequence of intermediate classes through which to route flow from~$\classt{}{T}$ to~$\classt{}{T'}$. See Figure~\ref{fig:mgraphflow}. We specified in our companion paper~\cite{eppfrishtri} how to route the flow through each intermediate class so that the congestion across each edges between a pair of classes is at most~$O(1)$, then made use of the existing flows within each class guaranteed by the inductive hypothesis to bound the resulting amount of flow within each class~$\classt{}{T}$. We showed that the latter is at most $O(1)\cdot \rho_T$, giving overall congestion~$O(1)^l$, where $l$ is the number of induction levels. Since~$X$ is a balanced separator we have $l = O(\log n)$; the lemma now follows from Lemma~\ref{lem:expmixing}.
\end{proofsketch}
The full proof of Lemma~\ref{lem:condexp} is in our companion paper~\cite{eppfrishtri}. We will use the phrase ``non-hierarchical framework'' to describe this set of conditions\textemdash which apply to the chains we study when the underlying graph $G$ has bounded treewidth and degree. Although Jerrum, Son, Tetali, and Vigoda~\cite{jerrumprojres} did not consider bounded-treewidth graphs generally, these conditions do allow their projection-restriction technique to be applied. In effect, Lemma~\ref{lem:condexp} and its proof, which we gave in our previous work, characterize a sufficient set of conditions for applying Jerrum, Son, Tetali, and Vigoda's technique: specifically, one can, instead of routing flow internally through each intermediate class, simply treat the construction above as a flow in the projection graph, concluding that the projection graph has a good spectral gap\textemdash then apply~\cite{jerrumprojres}. The first main technical contribution of this paper is in Section~\ref{sec:altfwintro}, in which we give an alternative set of conditions\textemdash which we will call our ``hierarchical framework''\textemdash that allows us to handle underlying graphs of unbounded degree (though treewidth still must be bounded), and to handle chains other than the hardcore model. This will allow us to complete the proofs of Theorems~\ref{thm:indsetmix}, \ref{thm:lcolor}, and~\ref{thm:domsetmix}.

\section{$\lambda = 1$: Unbounded degree}
\label{sec:altfwintro}
\subsection{Hierarchical framework}
We now sketch ``hierarchical'' framework conditions that guarantee rapid mixing in the case of unbounded degree (when treewidth is bounded). Several of the chains we consider satisfy these conditions so long as the treewidth of the underlying graph is bounded. This is the first main technical contribution in this paper. In the original framework, we assumed that the classes were approximately the same size. Although all of the graphs to which we apply this hierarchical framework satisfy this condition in graphs with bounded treewidth and degree, this is not the case when the degree is unbounded. Fortunately, in the case of independent sets, partial $q$-colorings, dominating sets, and $b$-edge covers, we solve this problem with some modifications to the framework. 

\subsection{Independent sets}
\begin{figure}
\centering
\includegraphics[width=15em]{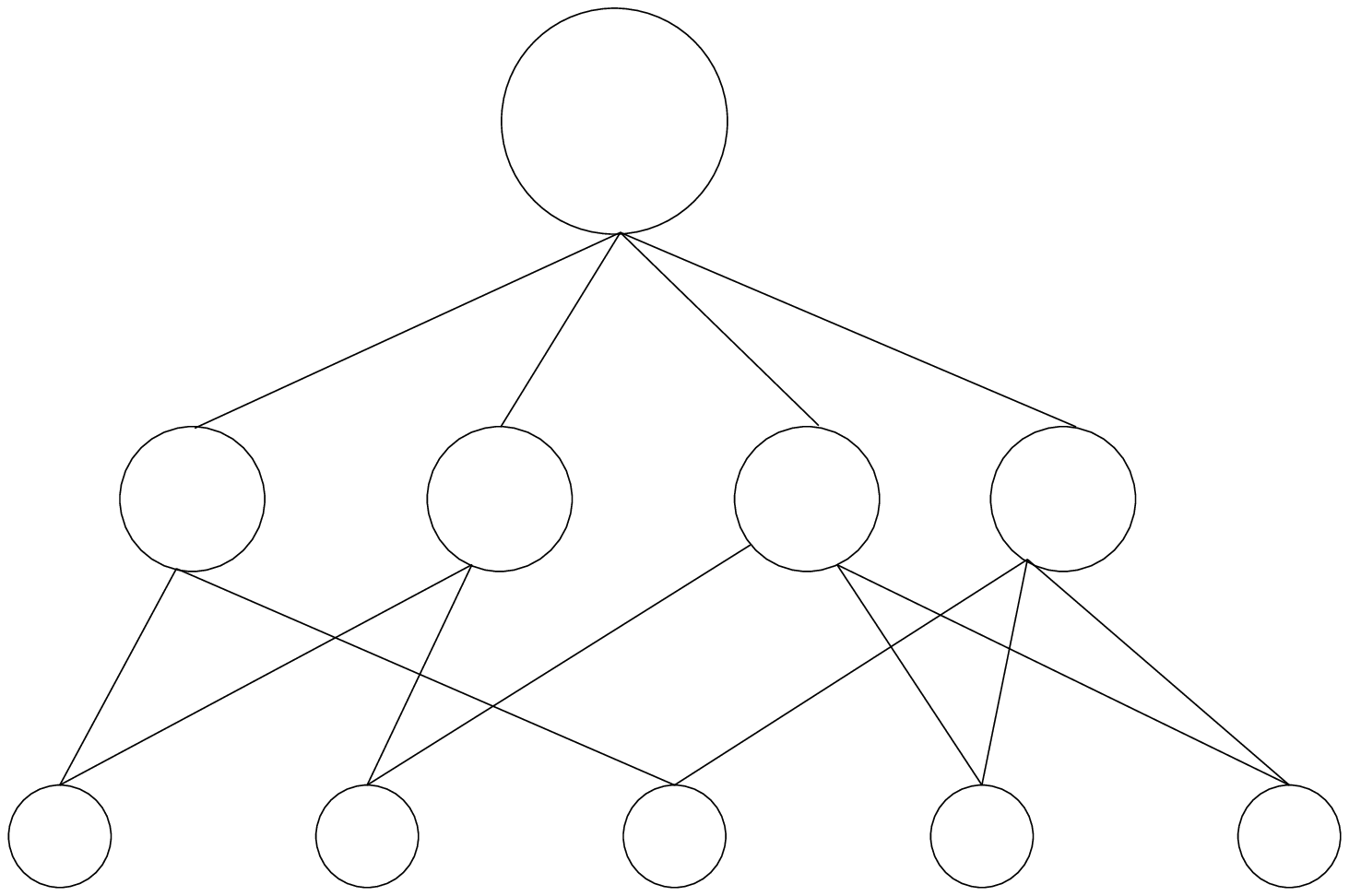}
\hspace{3cm}
\includegraphics[width=8em]{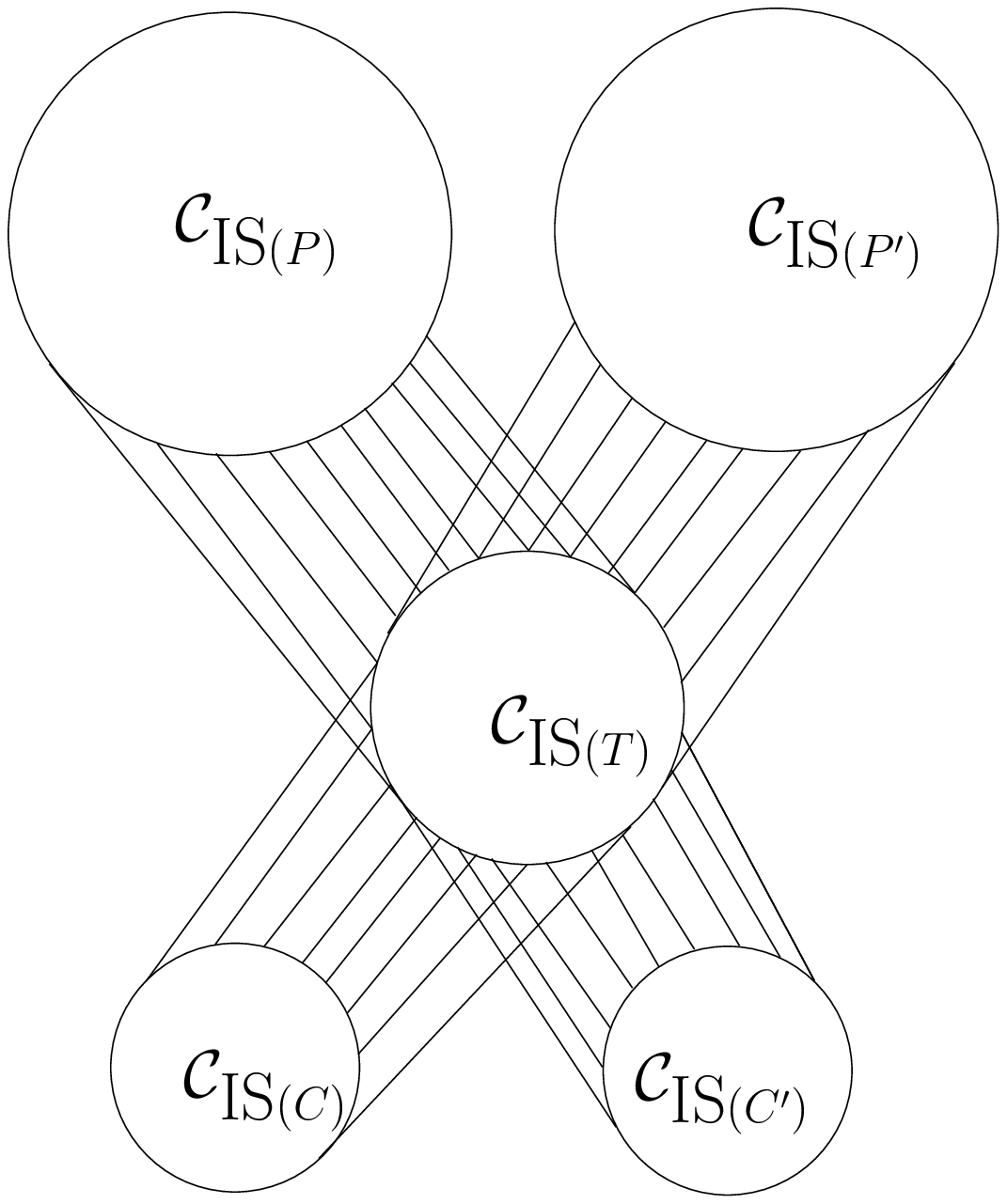}
\caption{Left: a schematic representation of the classes in the independent set Glauber graph and edges between them when the degree is unbounded. Right: a class $\classt{IS}{T}$, with two parents, $\classt{IS}{P}$ and $\classt{IS}{P'}$, and two children, $\classt{IS}{C}$ and $\classt{IS}{C'}$. (Classes with larger cardinality are drawn larger.) The parallel edges depict the fact that a child class always has every one of its vertices adjacent to a vertex in a given parent class, and that the edges between any given pair of classes are vertex-disjoint.}
\label{fig:ismeta}
\end{figure}

In the proof of Lemma~\ref{lem:condexp}, the assumption that the classes were approximately the same size allowed every class~$\classt{IS}{T}$ to route flow for all pairs of vertices without being too congested, because $\classt{IS}{T}$ is sufficiently large. Once we discard this assumption, we need to specify explicitly the path through which a given $\classt{IS}{T}$ routes flow to each~$\classt{IS}{T'}$. We construct a flow in which for every such $\classt{IS}{T}, \classt{IS}{T'}$, every intermediate class $\classt{IS}{T''}$ that handles flow between sets $S \in \classt{IS}{T}$ and $S' \in \classt{IS}{T'}$ is larger than one of $\classt{IS}{T}$ or $\classt{IS}{T'}$. We then bound the number of pairs of sets, relative to $|\classt{IS}{T''}|$, for which $\classt{IS}{T''}$ carries flow.

To accomplish this, we observe that for any~$\classt{IS}{T}, \classt{IS}{T'}$ such that there exists \emph{one} move between an independent set in $\classt{IS}{T'}$ and an independent set in $\classt{IS}{T}$, either \emph{every} independent set in $\classt{IS}{T'}$ has a move to some independent set in $\classt{IS}{T}$, or vice versa. This move consists of dropping some vertex $v$ from $T' \subseteq X$ to obtain $T$, i.e.~$T = T' \setminus \{v\}$. We call $\classt{IS}{T}$ a \emph{parent} of $\classt{IS}{T'}$, and $\classt{IS}{T'}$ a \emph{child} of $\classt{IS}{T}$. See Figure~\ref{fig:ismeta}. Since the set of edges connecting vertices in~$\classt{IS}{T}$ with vertices in~$\classt{IS}{T'}$ forms a matching, this implies that $|\classt{IS}{T}| \geq |\classt{IS}{T'}|$. In fact, whenever $T \subseteq T'$, we have $|\classt{IS}{T}| \geq |\classt{IS}{T'}|$. We route flow between any pair of classes $\classt{IS}{T}$ and $\classt{IS}{T'}$ along a path through a ``least common ancestor''. Since for every class~$\classt{IS}{T''}$ on this path, either $|\classt{IS}{T''}| \geq |\classt{IS}{T}|$ or $|\classt{IS}{T''}| \geq |\classt{IS}{T'}|$, we obtain a bound on congestion that we make precise in Appendix~\ref{sec:mainfwcond}.

In the proof sketch of Lemma~\ref{lem:condexp} (Section~\ref{sec:iscond}), for every pair $S \in \classt{}{T}, S' \in \classt{}{T'} \neq \classt{}{T}$, we found a sequence of classes $\classt{}{T} = \classt{}{T_1}, \classt{}{T_2}, \dots, \classt{}{T_{k-1}}, \classt{}{T_k} = \classt{}{T'}$ through which to route the $S-S'$ flow. As discussed in Section~\ref{sec:altfwintro}, when the degree is unbounded, the classes are no longer nearly the same size\textemdash so if this sequence is chosen carelessly, some~$\classt{}{T_i}$ may carry flow for too many $S-S'$ pairs. We therefore choose the sequences carefully: the parent-child relationships induce a partial order $\prec$ on the classes with a unique maximal element, where $\classt{}{T} \succ \classt{}{T'}$ implies $|\classt{}{T}| \geq |\classt{}{T'}|$. We choose our sequence so that for some $i$ with $1 \leq i \leq k$, $|\classt{}{T_1}| \leq |\classt{}{T_2}| \leq \cdots \leq |\classt{}{T_i}| \geq |\classt{}{T_{i+1}}| \geq \cdots |\classt{}{T_{k-1}}| \geq |\classt{}{T_k}|$. 

\subsection{Hierarchical Framework Conditions}
\label{sec:mainfwcondlist}
The conditions are as follows. Conditions~\ref{maincond2} through~\ref{maincond4} are new and concern the partial order described above; Condition~\ref{maincond1} and Conditions~\ref{maincond5} through~\ref{maincond7} are as in Section~\ref{sec:fwcond}.
\begin{enumerate}
\item \label{maincond1} The vertices of $\flipg{}{G}$ can be partitioned into a set $\mathcal{S}$ of classes, where $|\mathcal{S}| = O(1).$
\item \label{maincond2} There exists a partial order $\prec$ on the classes in $\mathcal{S}$, such that whenever $\classt{}{T}, \classt{}{T'} \in \mathcal{S}$ and $\classt{}{T} \succ \classt{}{T'}$, we have $|\classt{}{T}| \geq |\classt{}{T'}|$.
\item \label{maincond3} The partial order $\prec$ has a unique maximal element.
\item \label{maincond4} Whenever an edge exists between vertices in $\classt{}{T}$ and $\classt{}{T'}$ with $\classt{}{T} \succ \classt{}{T'}$, the number of such edges is $|\classt{}{T'}|$. 
\item \label{maincond5} For every pair of classes $\classt{}{T}$ and $\classt{}{T'}$ that share an edge, the maximum degree, in $\classt{}{T}$, of a vertex in $\classt{}{T'}$, is $O(1)$, and the maximum degree, in $\classt{}{T'}$, of a vertex in $\classt{}{T}$, is $O(1)$.
\item \label{maincond6} Each class in $\mathcal{S}$ is the Cartesian product of two graphs ${\flipg{}{G_1}}$ and $\flipg{}{G_2}$, each of which can be recursively partitioned in the same way as $\flipg{}{G}$.
\item \label{maincond7} The recursive partitioning mentioned in Condition~\ref{maincond6} reaches the base case (graphs with one or zero vertices) in $O(\log n)$ levels of recursion.
\end{enumerate}

\begin{restatable}{lemma}{lemaltcondexp}
\label{lem:altcondexp}
Given a graph $\flipg{}{G}$ satisfying the conditions in Section~\ref{sec:mainfwcondlist}, the expansion of~$\flipg{}{G}$ is $\Omega(1/n^c)$, where $c = O(1)$.
\end{restatable}
We defer the proof of Lemma~\ref{lem:altcondexp} to Appendix~\ref{sec:mainfwcond}.

\appendix

\section{Discussion of method and open questions}

\subsection{Application to graphical models}
Prior work \cite{bordkang, ge2012graph} has shown that related chains, including \emph{softcore models}\textemdash in which the sampled sets need not be independent\textemdash mix rapidly on graphs of bounded treewidth. However, all of the Glauber dynamics we consider pertain to combinatorial sampling problems, in which one is sampling a subset of either the vertices or the edges of a graph, where the subsets must obey certain constraints, e.g. independence. As a result, and as Bordewich and Kang~\cite{bordkang} note, their technique does not extend to these models.

Similarly, in the setting of probabilistic graphical models, De Sa, Zhang, Olukotun, and R\'e~\cite{hierwidth} considered graphs with bounded \emph{hierarchy width}. They showed\textemdash via arguments similar to the projection-restriction technique~\cite{jerrumprojres}\textemdash that graphs with logarithmically bounded hierarchy width admit rapid mixing for the Glauber dynamics on models with bounded \emph{maximum factor weight}. It is straightforward to apply their argument to the \emph{Ising} and \emph{Potts} models with fixed parameters, on graphs of bounded treewidth and degree. This case of these models also admits application of projection-restriction (and in the special case of the path graph Jerrum, Sinclair, Tetali, and Vigoda observed this for the Ising model~\cite{jerrumprojres}), and it fits our framework. Since our framework does not give a substantial improvement on existing results for these models, we do not address them in detail in this paper; we simply note that the framework we developed in our prior work~\cite{eppfrishtri} applies to these cases and to every undirected graphical model having only pairwise and unary factors, bounded maximum factor weights, constantly many values for each random variable, and bounded treewidth and degree. This shows that the framework unifies these models\textemdash in which all states have positive probability and which prior work has addressed in these graphs\textemdash with combinatorial chains where some states have zero probability\textemdash for which our results are new. We give a brief sketch of how to apply our framework in Appendix~\ref{sec:isingpotts}. See Bordewich, Greenhill, and Patel~\cite{bgpatel} and Chen, Liu, and Vigoda~\cite{chen2020rapiduniq} for definitions of and results for these models.

\subsection{Comparison with the projection-restriction technique}
\label{sec:projres}
In the special cases of the hardcore model, $q$-colorings, and partial $q$-colorings\textemdash as well as the bounded-treewidth-and-degree case of $b$-edge covers and $b$-matchings, one could reframe our inductive step in terms of the projection-restriction technique of Jerrum, Son, Tetali, and Vigoda~\cite{jerrumprojres}. Furthermore, as we have noted, Heinrich~\cite{heinrich2020glauber} used the projection-restriction technique for $q$-colorings.

Indeed, the initial framework conditions we summarized in Section~\ref{sec:fwcond} are seen to imply rapid mixing\textemdash provided an additional mild condition is satisfied\textemdash using the projection-restriction technique, as we observed in our companion paper. The idea is to first partition the Glauber graph into classes, each of which is a Cartesian product of smaller Glauber graphs over an underlying graph at most half the size of the original graph\textemdash just as we have done. One then defines a \emph{projection} Markov chain whose states are identified with the classes, where each class has probability mass proportional to its cardinality. The transition probabilities between classes are then proportional to the numbers of edges between classes (up to normalization by the degree of each boundary vertex).

If one then finds a good upper bound on the mixing times of both the projection chain and each of the individual class chains, one can derive a good upper bound on the mixing time of the original chain. One then needs either to incur an $O(1)$ congestion factor at each inductive step of the decomposition, combined with an $O(\log n)$ induction depth\textemdash or alternatively to bound a particular quantity $\gamma$ that describes, essentially, the number of edges between any given class and the rest of the graph.

To find such an upper bound, one could naturally view our ``classes'' as individual states in the projection chain, with probability masses proportional to their cardinalities. Our construction of flow between classes would become, under this view, a multicommodity flow in the projection chain.

However, as discussed in the introduction, constructing the flow in the projection chain still requires showing that each chain satisfies the framework conditions\textemdash and in the case of the hardcore model and dominating sets, one needs the hierarchical framework conditions. Furthermore, as formulated by Jerrum, Son, Tetali, and Vigoda~\cite{jerrumprojres}, the projection-restriction technique requires that the decomposition be a partition of the state space. By contrast, we deal with several chains in which the ``classes'' overlap or are internally not Cartesian products: namely dominating sets, $b$-edge covers (when the degree is unbounded), maximal independent sets, and maximal $b$-matchings. For these chains, our flow analysis provides a finer mechanism for dealing with these structural challenges.

Thus we present our construction purely in combinatorial terms\textemdash as opposed to considering a flow in a projection state space\textemdash for two key reasons: (i) to deal with non-independence as in the third and fourth of our main theorems, and (ii) because we believe our construction makes the graphical analysis of these chains more intuitive.

\subsection{Further discussion of prior work}
\label{sec:otherprior}
Sly~\cite{sly} showed that, except for restricted values of $\lambda < 1$, the hardcore model approximate sampling problem (obtaining an FPRAS) is hard (and thus the Glauber dynamics does not mix rapidly) on general graphs unless RP = NP. More precisely, Sly showed hardness when $\lambda_c < \lambda < \lambda_c + \varepsilon$ where~$\lambda_c$ is a certain \emph{uniqueness threshold} that depends on the degree of the input graph. Sly and Sun~\cite{slysun} and independently Galanis, {\v{S}}tefankovi{\v{c}}, and Vigoda~\cite{gsvhardness} later proved hardness for all $\lambda > \lambda_c$.  On the other hand, Anari, Liu, and Oveis Gharan \cite{anarimixis} used a technique known as \emph{spectral independence} to obtain rapid mixing for the hardcore Glauber dynamics when $\lambda$ is below the uniqueness threshold. They showed, by exhibiting an infinite family of examples, that the technique they used could not be further improved (namely beyond the uniqueness threshold) even for trees. By contrast, we show that rapid mixing, for all fixed values of $\lambda$, indeed holds not only for trees but for all graphs of bounded treewidth. Chen, Liu, and Vigoda~\cite{clvopt} showed $O(n\log n)$ mixing for bounded-degree graphs when $\lambda < \lambda_c - \varepsilon$ for every~$\varepsilon > 0$, with a dependence on~$\varepsilon$ in the mixing time. Chen, Galanis, {\v{S}}tefankovi{\v{c}}, and Vigoda \cite{chen2020rapid} generalized the technique of Anari, Liu, and Oveis Gharan to all 2-spin systems; Feng, Guo, Yin, and Zhang \cite{feng2020rapid} applied it to graph colorings.

Other results exist for trees beyond the uniqueness threshold, however: Martinelli, Sinclair, and Weitz~\cite{martinellitrees} showed that the dynamics on $q$-colorings ($q \geq \Delta + 2$) mixes in $O(n\log n)$ time on the complete $\Delta - 1$-ary tree with $n$ nodes.  Lucier, Molloy, and Peres~\cite{luciercolorings} showed that the dynamics mixes rapidly on general trees of bounded degree, namely in time $O(n^{O(1+\Delta/(q\log \Delta))})$. Restrepo, {\v{S}}tefankovi{\v{c}}, Vigoda, and Yang~\cite{slowtrees} showed that for certain trees the mixing time slows down when~$\lambda$ is sufficiently large.

Prior work also exists for $q$-colorings of bounded-treewidth graphs: Berger, Kenyon, Mossel, and Peres~\cite{bkmp} showed rapid mixing for $q$-colorings of regular trees with unbounded degree. Tetali, Vera, Vigoda, and Yang~\cite{tvvy} gave upper and lower bounds for complete trees. Vardi~\cite{varditw} showed that the so-called \emph{single-flaw} dynamics\textemdash a variaton on the Glauber dynamics in which at most one monochromatic edge is permitted in a valid state\textemdash mixes rapidly on bounded-treewidth graphs when $q \geq (1+\varepsilon)\Delta$, for any fixed parameter $\varepsilon > 0$. The proof used the vertex separaton properties of bounded-treewidth graphs to construct a multicommodity flow with bounded congestion, although the construction was substantally different from our divide-and-conquer approach. Dyer, Goldberg, and Jerrum~\cite{dyergj} showed rapid mixing when the degree of the graph is at least $2t$ and $q \geq 4t$, where $t$ is the treewidth. On the other hand, Heinrich~\cite{heinrich2020glauber} showed that the Glauber dynamics on $q$-colorings of a bounded-treewidth graph mixes rapidly when $q \geq \Delta + 2$. Our construction, as we will discuss in more detail in Section~\ref{sec:priorsim}, bears some similarity to Heinrich's. We also require that $q$ (and therefore $\Delta$) be bounded. However, due to a more general analysis of the state spaces of Glauber dynamics as graphs, we obtain a more general framework that holds for a greater variety of chains. We obtain an improvement over~\cite{dyergj} and~\cite{heinrich2020glauber} in the dependence on treewidth when $\Delta < 2t$ or when $q < 4t$ and $\Delta < t^2$. We additionally show that the Glauber dynamics on the \emph{partial} $q$-colorings of $G$ mix rapidly for all fixed $\lambda > 0$ when $q \geq \Delta + 2$ is bounded.

Planar graphs have unbounded but sublinear treewidth. For planar graphs, Hayes~\cite{hayestw} showed that the Glauber dynamics on $q$-colorings of a planar graph of maximum degree $\Delta$ mixes rapidly when $q \geq \Delta + O(\sqrt{\Delta})$. Later, Hayes, Vera, and Vigoda~\cite{hvvcolorings} proved rapid mixing for $q$-colorings of planar graphs when $q = \Omega(\Delta/\log \Delta)$, generalizing further to a spectral condition on the adjacency matrix of the graph.

Bez{\'a}kov{\'a} and Sun showed \cite{ischordal} that the hardcore model mixes rapidly in chordal graphs with bounded-size separators. Lastly, Chen, Galanis, {\v{S}}tefankovi{\v{c}}, and Vigoda applied the spectral independence technique to prove that the Glauber dynamics on the $q$-colorings of a triangle-free graph with dgree $\Delta$ mixes rapidly provided that $q \geq \alpha\Delta + 1$, where $\alpha$ is greater than a threshold approximately equal to $1.763$. We show that when the treewidth and degree of $G$ are bounded, $G$ need not be triangle free, and it suffices that $q \geq \Delta + 2$ be bounded. We prove a similar result for the natural Glauber dynamics on \emph{partial} $q$-colorings.

Although our mixing results are new, Wan, Tu, Zhang, and Li showed \cite{wancounttw} that exact counting of independent sets is fixed-parameter tractable in treewidth. Furthermore, our result does not technically constitute a proof of fixed-parameter tractability, as the treewidth appears in the exponent of the polynomial we obtain. For this problem and all the other problems we consider, the problem of exact counting\textemdash and therefore also uniform sampling\textemdash has already been solved on the graphs we consider by an extension of Courcelle's theorem \cite{reinhardmsocount}. In fact, the standard reduction from approximate sampling to approximate counting \cite{sinclairsampling} gives a somewhat different rapidly mixing Markov chain on a larger state space. Nonetheless, our result does settle the question of rapid mixing for a natural chain, and it implies a simpler scheme for approximately sampling independent sets than one would obtain via this reduction.

Such a scheme is known as a \emph{fully polynomial randomized approximation scheme (FPRAS)}. Huang, Lu, and Zhang provided an FPRAS for sampling $b$-edge covers in general graphs when $b \leq 2$, and for sampling $b$-matchings when $b \leq 7$ \cite{huang2016canonical}. This FPRAS relied on a rapid mixing argument for a somewhat different Markov chain than ours. Existing dominating set results for certain regular graphs are also known \cite{bezakova2019approximation}.

Exact counting of maximal independent sets\textemdash which would give an FPRAS by the equivalence of counting and sampling\textemdash was shown in \cite{exactchordal} to be hard for chordal graphs but is known \cite{chenminimalfpt} to be tractable in graphs of bounded treewidth. However, again our result improves on the simplicity of existing algorithms.

Our approach is also inspired by Kaibel's~\cite{kaibelexp} construction of a flow with bounded congestion in any graph whose vertices are hypercube vertices and whose edges can be partitioned into bipartite graphs in a hierarchical fashion.

\subsection{Open questions}
\label{sec:conclusion}
We have developed a framework whose application shows rapid mixing for several natural chains on combinatorial structures in graphs of bounded treewidth. However, some work is required in showing that each of the structures satisfies the conditions of the framework. We hope that a more robust version of the framework can be developed that further unifies these techniques.

In particular, all of the structures we have analyzed satisfy the conditions of Courcelle's theorem, as noted previously. It would be interesting to determine whether the framework can be extended to all structures satisfying these conditions.

The fact that our results hold for all values of $\lambda > 0$ is not especially surprising, as Ioannis Panageas has observed, since the limiting case $\lambda = \infty$ corresponds to the optimization version of each problem, and the case $\lambda = 1$ corresponds to uniform sampling; as stated in the introduction, both of these problems are already known to be fixed-parameter tractable in treewidth. (In fact, as we noted in the introduction, the extension of Courcelle's theorem, combined with the reduction from sampling to counting, applies to all values of $\lambda > 0$.) Nonetheless, our result does settle a missing case of the mixing question in some generality, through purely combinatorial methods.

In all of our mixing bounds, the dependency on the parameters\textemdash treewidth and degree\textemdash is bad. It would be interesting to see whether some refinement of our methods could give a truly fixed-parameter tractable result, in which the treewidth and degree do not appear in the exponent of~$n$.

\section{Carving width}
\label{sec:cw}
Since the present paper concerns bounded-treewidth and bounded-degree graphs, we note the equivalence between these properties and a parameter known as \emph{carving width.} The carving width of a graph is a density parameter that is weaker than treewidth, in the sense that high treewidth implies high carving width, but the converse is not true. Carving width is defined with respect to a so-called \emph{carving decomposition}~\cite{eppwidth} of a given graph $G$\textemdash in short, a binary tree $T$ whose leaves are identified with the vertices of $G$. Each node $X \in T$ is identified with the subgraph of $G$ induced by the vertices of $G$ (leaves of $T$) having $X$ as an ancestor in $T$. Each edge of $T$ induces a cut in $T$; this cut induces a partition of the leaves of $T$ (vertices of $G$) into two sets. This partition is naturally identified with a cut in $G$.

The \emph{width} of a carving decomposition is the maximum number of edges of $G$ across any such cut, where the maximum is taken over all edges in $T$. The \emph{carving width} of $G$ is the minimum width of a carving decomposition of $G$. See Seymour and Thomas~\cite{seymourcarving} for a detailed treatment. For our purposes, carving width is of interest due to its relationship to the treewidth and degree of a graph. Specifically, Eppstein~\cite{eppwidth} observed the following fact that follows from results of Nestoridis and Thilikos~\cite{sqrts} and of Robertson and Seymour~\cite{robcarving}:
\begin{lemma}
\label{lem:cwtwdeg}
Given a graph $G$ with maximum degree $\Delta$, let $\tw(G)$ denote the treewidth of $G$, and let $\cw(G)$ denote the carving width of $G$. For every graph $G$, $(2/3)(\tw(G)+1) \leq \cw(G) \leq \Delta(\tw(G)-1)$.
\end{lemma}
It follows from the definition of carving width that every graph with bounded carving width also has bounded degree. Combining this fact with Lemma~\ref{lem:cwtwdeg} implies the following:
\begin{corollary}
\label{cor:cwtwdeg}
A graph has bounded degree and treewidth if and only if it has bounded carving width.
\end{corollary}

\section{Additional Chain Definitions}
\label{sec:moredefs}
\subsection{Dominating sets, maximal independent sets, $b$-matchings, and $b$-edge covers}
\label{sec:dombmax}
\begin{definition}
\label{def:domset}
A \emph{dominating set} in a graph $G = (V, E)$ is a set $S \subseteq V$ of vertices such that for every vertex $v \in V$, either $v \in S$ or there exists some vertex $u \in S$ such that $(u, v) \in E$.
\end{definition}

$b$-matchings \cite{bmat} and $b$-edge covers \cite{khanpothenbedge, Khan2016AN3} generalize the definitions of matchings and edge covers respectively:
\begin{definition}
\label{def:bmatching}
Let $G = (V, E)$ be a graph. Let $b: V \rightarrow \mathbb{Z}^{\geq 0}$ be any function assigning a nonnegative integer to each vertex. A \emph{$b$-matching} in a graph $G = (V, E)$ is a set $S \subseteq E$ of edges such that every $v \in V$ has at most $b(v)$ incident edges in $S$.
\end{definition}

\begin{definition}
\label{def:bedgecover}
Let $G = (V, E)$ be a graph. Let $b: V \rightarrow \mathbb{Z}^{\geq 0}$ be any function assigning a nonnegative integer to each vertex. A \emph{$b$-edge cover} in a graph $G = (V, E)$ is a set $S \subseteq E$ of edges such that every $v \in V$ has at least $b(v)$ incident edges in $S$.
\end{definition}

Sometimes, as in the result by Huang, Lu, and Zhang \cite{huang2016canonical}, $b$-edge covers and $b$-matchings are defined so that $b$ is a constant, i.e. $b(u) = b(v)$ for all $u, v \in V$.

For dominating sets, $b$-edge covers, and $b$-matchings, we consider a chain similar to the hardcore dynamis in Definition~\ref{def:hardcore}, except that in the case of $b$-edge covers and $b$-matchings, we are of course selecting edges instead of vertices. Also, in the case of dominating sets and $b$-edge covers, instead of verifying independence before \emph{adding} a vertex (or edge), we verify validity of a set (e.g. domination) before \emph{dropping} a vertex (or edge).

We also consider the Glauber dynamics on \emph{$q$-colorings}:
\begin{definition}
A \emph{$q$-coloring} of a graph $G$ is an assignment of a color from the list $[q] = \{1, 2, \dots, q\}$ to each vertex of $G$, such that no two adjacent vertices have the same color.
\end{definition}
\begin{definition}
A \emph{partial $q$-coloring} of a graph $G$ is an assignment of a color from $[q]$ to each of a subset of the vertices of $G$, such that no two adjacent vertices have the same color.
\end{definition}

The Glauber dynamics on the partial $q$-colorings of $G$ is as follows:
\begin{definition}
\label{def:partqglauber}
Let the Glauber dynamics on the partial $q$-colorings of a graph $G$ be the following chain defined with respect to $\lambda > 0$:
\begin{enumerate}
\item Let $X_0$ be an arbitrary partial $q$-coloring of $G$.
\item For $t \geq 0$, select a vertex $v \in V(G)$ uniformly at random, and select a color $c \in [q+1]$ uniformly at random.
\item If $c = q+1$, then:
\subitem If $v$ is already colored in $X_t$, remove the coloring of $v$ with probability $1/(\lambda+1)$.
\subitem Otherwise, let $X_{t+1} = X_t$.
\item If $c \leq q$, then:
\subitem If $v$ is \emph{not} already colored with $c$ in $X_t$, set the color of $v$ to $c$ with probability $\lambda/(\lambda+1)$.
\subitem Otherwise, let $X_{t+1} = X_t$.
\end{enumerate}
\end{definition}

Finally, the Glauber dynamics on the (complete) $q$-colorings of $G$ is as follows (for this chain we do not define a biased version):
\begin{definition}
\label{def:qglauber}
Let the Glauber dynamics on the $q$-colorings of a graph $G$ be the following chain:
\begin{enumerate}
\item Let $X_0$ be an arbitrary $q$-coloring of $G$.
\item For $t \geq 0$, select a vertex $v \in V(G)$ uniformly at random, and select a color $c \in [q]$\textemdash other than the color of $v$\textemdash uniformly at random.
\item If $v$ has no neighbor with color $c$, then change the color of $v$ to $c$ with probability 1/2 to obtain $X_{t+1}$.
\item Otherwise, do nothing, i.e. let $X_{t+1} = X_t$.
\end{enumerate}
\end{definition}

\begin{figure}
\centering
\includegraphics[width=15em]{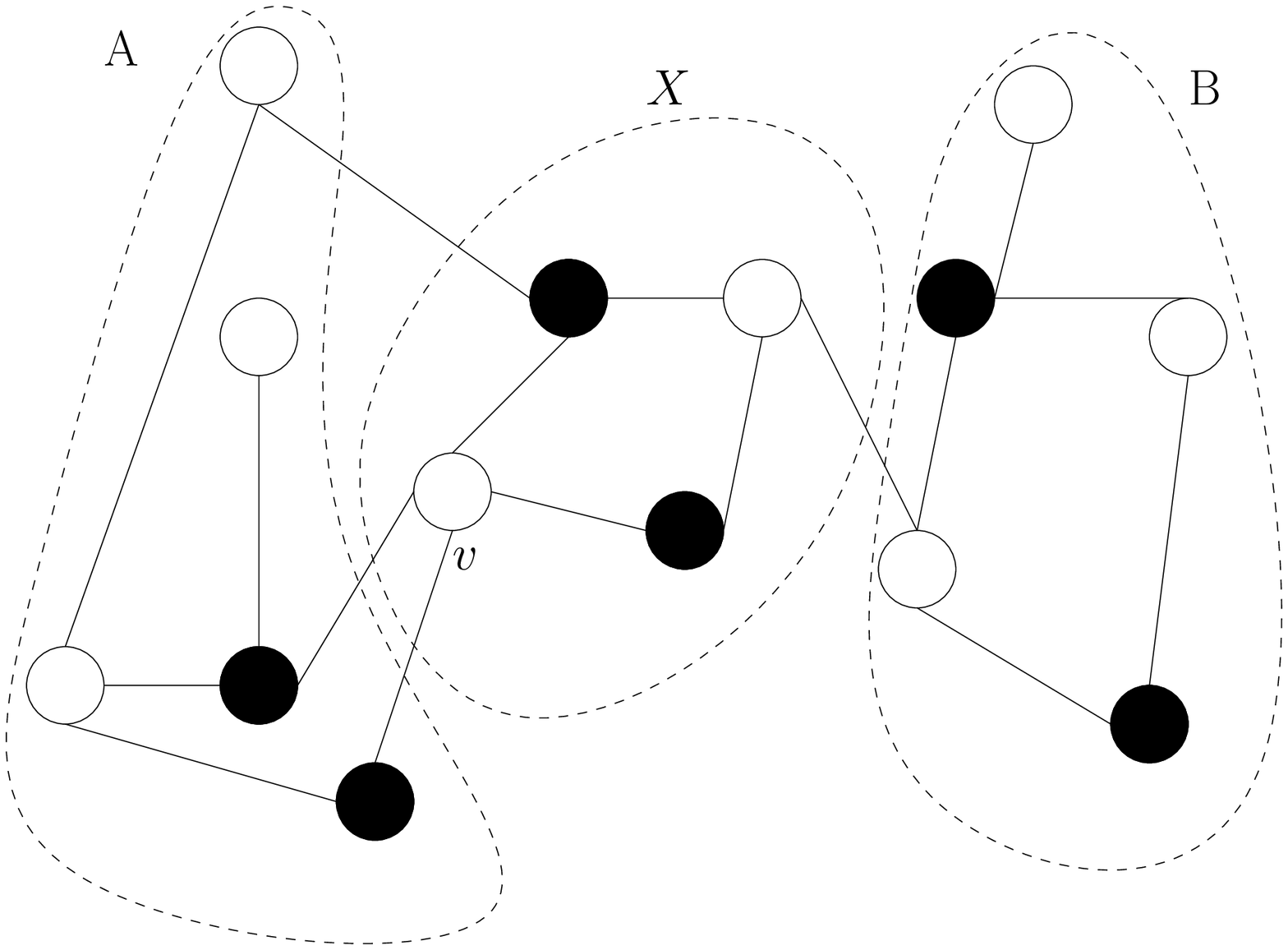}\hspace{1cm}
\includegraphics[width=15em]{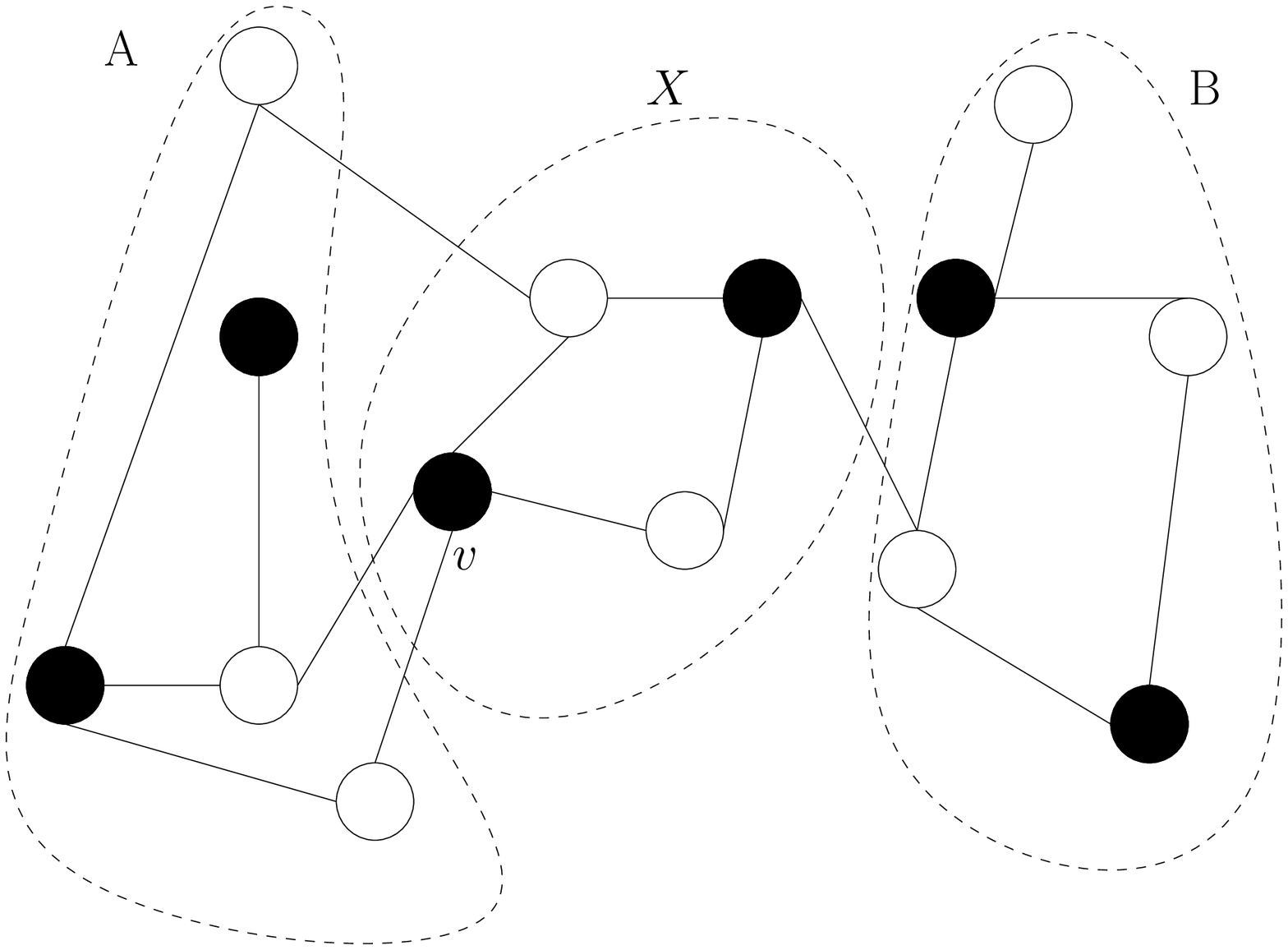}
\caption{Two maximal independent sets in a graph $G$: $S$ (left) and $S'$ (right). $S$ and $S'$ differ by a move, with the separator $X$ inducing the classes to which the sets belong. $S'$ results from adding $v$ to $S$, removing the neighbors of $v$, and adding some of the neighbors of the removed vertices.}
\label{fig:misflip}
\end{figure}

We define a graph whose vertices are the maximal independent sets of an underlying graph $G$, and then define the chain as a random walk on this graph:
\begin{definition}
\label{def:misgraph}
Given a graph $G = (V, E)$, let the \emph{maximal independent set Glauber graph} be the graph $\flipg{MIS}{G}$ whose vertices are the maximal independent sets of $G$, and whose edges are the pairs of maximal independent sets that differ by one \emph{move}, where a move is defined as:
\begin{enumerate}
\item adding one vertex $v$ to a given independent set $S \subseteq V$,
\item removing every $u \in S$ such that $(u, v) \in E$, and
\item adding a subset of the vertices at distance two in $G$ from $v$.
\end{enumerate}
\end{definition}
Since $\flipg{MIS}{G}$ is undirected, we also define the reversal of a move as a move. See Figure~\ref{fig:misflip} for an example of a move.

\begin{restatable}{lemma}{lemmisgraphconn}
\label{lem:misgraphconn}
The graph $\flipg{MIS}{G}$ in Definition~\ref{def:misgraph} is connected.
\end{restatable}
\begin{proof}
The proof relies on an easy greedy transformation argument and is in Appendix~\ref{sec:appdetails}.
\end{proof}

For maximal $b$-matchings, we define a Glauber graph similar to the maximal independent set Glauber graph, except that we are of course selecting edges instead of vertices in our sets. A move consists of adding some edge $e = (u, v)$ to the $b$-matching, then removing edges incident to $u$ and $v$ as needed until a valid $b$-matching is obtained, then adding edges incident to neighbors of $u$ and $v$ as needed to obtain maximality.

\section{Derivation of upper bounds in main theorems}
\label{sec:mixingder}
We now analyze the specific polynomial upper bounds that we obtain from each version of the framework.

In the following bounds, we consider all logarithms to be base two, unless otherwise stated. The $\log n$ terms in the exponents of these bounds come from the balanced separators guaranteed by bounded treewidth. Technically, as we have defined balanced separators, one of the two mutually disconnected subgraphs obtained by removing a balanced separator may have size greater than $n/2$. However, one can show~\cite{ericksontw} that no \emph{connected component} of the resulting disconnected graph has size greater than $n/2$. It is straightforward to modify many of our proofs to account for Cartesian products with multiple factor graphs, by iterating Lemma~\ref{lem:cartflow}. When this is not possible, we will explicitly state the base we use.

We proved the following in our companion paper~\cite{eppfrishtri}:

\begin{restatable}{lemma}{lemnonhierexact}
\label{lem:nonhierexact}
Suppose a Glauber graph $\flipg{}{G}$ satisfies the conditions of the non-hierarchical framework. Then the mixing time of the corresponding Glauber dynamics is
$$O((2N/\mathcal{E}_{\min}) + 1)^{2\log n}\cdot \Delta_{\mathcal{M}}^2\log N),$$
where $\mathcal{E}_{\min}$ is the size of the smallest edge set between adjacent classes.
\end{restatable}
\begin{proofsketch}
The idea of the proof, which we gave in our companion paper, is that there exists a uniform multicommodity flow in which no matching~$\edget{}{T}{T'}$ carries more than~$N^2$ un-normalized flow, since this is the total amount of flow to be exchanged in the uniform multicommodity flow. Therefore each vertex in each class~$\classt{}{T}$ brings in at most~$2N^2/\mathcal{E}_{\min}$ un-normalized flow across all of its edges out of~$\classt{}{T}$ combined. The result then follows from using the inductively defined flow within~$\classt{}{T}$ to route this flow, with the factor of 2 coming from accounting for both ``inbound'' and ``outbound'' flow, and with the~$2\log n$ exponent and the~$\Delta_{\mathcal{M}}^2\log N$ terms coming from the squaring that takes place in Lemma~\ref{lem:expmixing}.

The arXiv version we have cited gives a term~$4N/\mathcal{E}_{\min}$ instead of the~$2N/\mathcal{E}_{\min}$ term we have given here. A newer version of the companion paper has the latter bound. To obtain the improvement, it suffices to take the proof in the arXiv version~\cite{eppfrisharx}, Lemma 7.1, and observe that the incoming ``through'' flow described, combined with the inbound flow, can only sum to~$N^2/(|\edget{}{T}{T'}|$ units.
\end{proofsketch}

In the case of $q$-colorings, tracing the constant factors in the proof of Lemma~\ref{lem:colexp}, we see that~$N/\mathcal{E}_{\min} \leq q^{2\Delta(t+1)},$ that $\Delta_{\mathcal{M}}\leq (q-1)n$, and that $N \leq q^n$. Combining this with Lemma~\ref{lem:expmixing} and Lemma~\ref{lem:flowexp} gives the bound claimed in Theorem~\ref{thm:lcolor}. More precisely, the bound is
$$((2(q^{2\Delta(t+1)})+1)^{2\log n}) = O((q-1)^2\log q\cdot n^{4(t+1)\Delta\log q+7}).$$

We will prove the following in Appendix~\ref{sec:mainfwcond}:
\begin{restatable}{lemma}{lemhierexact}
\label{lem:hierexact}
Suppose a Glauber graph $\flipg{}{G}$ satisfies the conditions of the hierarchical framework. Then the mixing time of the corresponding Glauber dynamics is
$$O(((2K+1)^{2\log n})\cdot \Delta_{\mathcal{M}}^2\log N),$$
where $\Delta_{\mathcal{M}}$ is the maximum degree of the Glauber graph $\flipg{}{G}$, $n = |V(G)|$, $K$ is the number of classes in the partition, and $N = |V(\flipg{}{G})|$.
\end{restatable}

The unbiased case of the bound in Theorem~\ref{thm:indsetmix} now follows from combining Lemma~\ref{lem:hierexact} with the observation that for this chain, $\log N \leq n$, $\Delta_{\mathcal{M}} = n$, and $K \leq 2^{t+1}$. Similarly, as we will see, the bound for the unbiased case of partial $q$-colorings in Theorem~\ref{thm:lcolor} will follow from the fact that $K \leq (q+1)^{t+1}$, $\Delta_{\mathcal{M}} \leq qn$, and $N \leq (q+1)^n$, so that
$$(2(K+1))^{2\log n}\Delta_{\mathcal{M}}^2\log N = n^{2\log{2(K+1)}}\Delta_{\mathcal{M}}^2\log N$$
$$\leq q^2\log (q+1)\cdot n^{2(t+2)\log (q+1)+5}.$$

\section{Bounded treewidth and degree: application of framework beyond independent sets}
\subsection{$q$-colorings}
\label{sec:nonhierqcol}
We now apply the non-hierarchical framework to $q$-colorings in graphs of bounded treewidth and degree. For reasons that will soon become apparent, we need to generalize to \emph{list colorings}:
\begin{definition}
\label{def:lcol}
A \emph{list coloring} of a graph $G = (V, E)$, given a function $L:V \rightarrow 2^{[q]}$ assigning a list of colors to each vertex in $V$, is a coloring of $G$ consistent with $L$. A \emph{partial list coloring} is a coloring of some of the vertices of $G$ consistent with $L$.
\end{definition}

We consider the Glauber graph $\flipg{COL}{G,L}$, defined as follows:
\begin{definition}
Let the Glauber graph $\flipg{COL}{G,L}$, given an input graph $G$ and a set of colors~$[q]$ and a function~$L$ as in Definition~\ref{def:lcol}, be the graph whose vertices are the list colorings of~$G$ consistent with~$L$, and whose edges are the pairs of list colorings~$C, C'$ that differ by a color assignment to exactly one vertex~$v \in V(G)$.
\end{definition}

The Glauber dynamics in Definition~\ref{def:qglauber} is the natural random walk on $\flipg{COL}{G,L}$, with self-loops added in the standard fashion. The following lemma therefore suffices to prove the first claim in Theorem~\ref{thm:lcolor}:
\begin{lemma}
\label{lem:colexp}
$\flipg{COL}{G,L}$, defined over a graph $G$ and a list $L: V(G) \rightarrow 2^{[q]}$, with $L(v) \geq \delta(v) + 2$ for every $v \in V(G)$, satisfies the conditions of the non-hierarchical framework whenever $G$ has bounded treewidth and degree and $q$ is fixed.
\end{lemma}
\begin{proof}
We partition $V(\flipg{COL}{G,L})$ into classes induced by a small balanced separator $X$, where each class is identified with a list coloring $T$ of $G[X]$. This partitioning satisfies Condition~\ref{cond5} since each class $\classt{COL}{T}$ consists of the tuples of the form $(C_A, C_B)$, where $C_A$ is a valid list coloring of~$G[A]$, and~$C_B$ is a valid list coloring of~$G[B]$\textemdash with~$A$ and~$B$ being the mutually disconnected subsets of~$V(G)$ resulting from the removal of~$X$. Here, we adjust the list~$L(u)$ for each $u \in N_A(X) \cup N_B(X),$ removing from~$L(u)$ every color that is assigned to a neighbor of~$u$ in~$X$ under the coloring~$C$.

The subproblems on~$A$ and~$B$ are independent, and that a move within~$\classt{COL}{T}$ corresponds to a move within either~$A$ or~$B$ but not both. Furthermore, the condition that~$L(u)\geq\delta(u)+2$ is preserved even after~$L$ is modified, since every color removed from~$L(u)$ corresponds to a neighbor of~$u$ in~$X$\textemdash i.e. a neighbor that is not part of the subproblem on~$A$ or~$B$. Condition~\ref{cond5} follows.

Condition~\ref{cond1} follows from the fact that $|X|$ and $q$ are bounded. Condition~\ref{cond2} can be seen from the bounded treewidth and degree of $G$ by considering the following mapping $f:V(\flipg{COL}{G,L}) \rightarrow \classt{COL}{T}$ for any $T$: given a list coloring $C \in V(\flipg{COL}{G,L})$, let $C' = f(C)$ be the list coloring that (i) agrees with $T$ on its restriction to $X$, (ii) agrees with $C$ on its assignment of colors to all vertices having no neighbor in $X$, and (iii) is consistent with both (i) and (ii) on its assignment of colors to neighbors of vertices in $X$.

We can always satisfy (iii) because for each $u \in N_A(X) \cup N_B(X)$, we have $|L(u)| \geq \delta(u) + 2$. (There may be multiple list colorings satisfying (iii); resolve ambiguity in defining~$f(C)$ via an arbitrary ordering on the list colorings of~$G$.) Condition~\ref{cond4} follows from a similar mapping.

Condition~\ref{cond3} follows from the definition of a move; Condition~\ref{cond6} follows from the bounded treewidth and degree of $G$.
\end{proof}

\subsection{$b$-edge covers and $b$-matchings}
\label{sec:bprobs}
For $b$-edge covers and $b$-matchings, we now apply the non-hierarchical framework in graphs of bounded treewidth and degree. As with independent sets, dealing with unbounded degree in $b$-edge covers requires the hierarchical framework. 

\begin{lemma}
\label{lem:bproblemssimple}
Given an input graph $G$ of bounded treewidth and degree, the Glauber dynamics on $b$-matchings and on $b$-edge covers satisfy the conditions of the non-hierarchical framework, when the maximum value of the function $b$ is bounded.
\end{lemma}
\begin{proof}
The proof for $b$-edge covers is similar to the proof of Lemma~\ref{lem:iscondexp}, with the following modifications.

In defining a $b$-edge cover, we are selecting subsets of edges instead of vertices. Thus, to define our chain on $b$-edge covers, we modify the chain on independent sets in the natural way: dropping or adding edges instead of vertices. The corresponding Glauber graph $\flipg{BEC}{G}$ is connected, since every $b$-edge cover has a path in $\flipg{BEC}{G}$ to the trivial $b$-edge cover (where every edge is selected). We identify each class $\classt{BEC}{T}$ with the set $T$ of edges chosen incident to vertices in $X$. Since degree is bounded and $|X| \leq t$, there are $O(1)$ classes, satisfying Condition~\ref{cond1}. 

Given a class $\classt{BEC}{T}$, we pass recursively to subproblems on $A$ and $B$, where we update $b(v)$ for each $v \in A \cup B$ according to the number of edges in $T$ incident to $v$. That is, for each vertex $u$ selected in $T$, and for each edge $(u, v)$ with $v \in A$ (similarly $v \in B$), decrement $b(v)$ when passing to the subproblem on $A$ (similarly $B$). The choices made in the $A$ subproblem and the $B$ subproblem are independent, giving the required Cartesian product structure for Condition~\ref{cond5}, and there are still $O(\log n)$ levels of recursion, satisfying Condition~\ref{cond6}. For Condition~\ref{cond3}, the proof is the same as for independent sets. Conditions~\ref{cond2} and~\ref{cond4} follow from a similar mapping argument to that in the proof of Lemma~\ref{lem:iscondclasssizes}.

The proof for $b$-matchings is similar to that for $b$-edge covers.
\end{proof}

\subsection{Maximal independent sets and maximal $b$-matchings}
The main idea of applying the framework to maximal independent sets and maximal $b$-matchings is similar to that for independent sets, $b$-matchings, and $b$-edge covers, but some adaptation is required: the definition of a move is somewhat different, and the proof that classes have the required Cartesian product structure has a few more details. We thus defer dealing with these chains to Appendix~\ref{sec:mismaxb}.

\section{Hierarchical framework}
\label{sec:mainfwcond}
In this section we complete the proof of the unbiased case of Theorem~\ref{thm:indsetmix} and Theorem~\ref{thm:lcolor}, by fully specifying the hierarchical framework, and showing that the chain on independent sets satisfies the conditions. Fully proving Theorem~\ref{thm:domsetmix} and Theorem~\ref{thm:mindsetmix} requires some adaptation of the framework, which we defer to Appendix~\ref{sec:adapting}.
\subsection{Proof that conditions of the hierarchical framework imply rapid mixing}
\label{sec:indsetindflow}
We are ready to prove the counterpart of Lemma~\ref{lem:condexp} for the hierarchical framework, from which the unbiased case of Theorem~\ref{thm:indsetmix} will follow. 

\lemaltcondexp*
\begin{proof}
We use the scheme in the proof of Lemma~\ref{lem:condexp}, with the following specification: when routing flow from $S \in \classt{}{T}$ to $S' \in \classt{}{T'} \neq \classt{}{T}$, we find a sequence of classes $\classt{}{T} = \classt{}{T_1}, \classt{}{T_2}, \dots, \classt{}{T_{k-1}}, \classt{}{T_k} = \classt{}{T'}$ as before, where each consecutive pair of classes in the sequence shares an edge in $\flipg{}{G}$. In the proof of Lemma~\ref{lem:condexp}, this sequence was arbitrary; we now require that, under the partial order $\prec$ in Condition~\ref{maincond2}, for some $1 \leq i\leq k$, $\classt{}{T_1} \prec \cdots \prec \classt{}{T_i} \succ \classt{}{T_{i+1}} \succ \cdots \succ \classt{}{T_k}$; Condition~\ref{maincond3} guarantees that this requirement can be satisfied.

We now bound the resulting congestion. As in the proof of Lemma~\ref{lem:condexp}, for $i = 2, \dots, k-1$, the congestion added to edges in $\classt{}{T_i}$ in the inductive step is at most $N^2/(|Y_i||Z_i|)\cdot c^{\log n - 1}$. Unfortunately, without assuming that the classes are approximately the same size, we can no longer say that $|Y_i| = \Omega(N)$ or $|Z_i| = \Omega(N)$. Instead, we argue as follows: thanks to the choice of our sequence, for every pair of classes $\classt{}{T}$ and $\classt{}{T'}$ that use a given class $T_i$ to route flow, either $|\classt{}{T_i}| \geq |\classt{}{T}|$ (and $|\classt{}{T_i}| \geq |\classt{}{T_{i-1}}|$) or $|\classt{}{T_i}| \geq |\classt{}{T'}|$ (and $|\classt{}{T_i}| \geq |\classt{}{T_{i+1}}|$). Assume the former case without loss of generality. For every pair of classes $\classt{}{T}$ and $\classt{}{T'}$ that use the edges between $\classt{}{T_{i-1}}$ and $\classt{}{T_i}$, $|\classt{}{T}| \leq |\classt{}{T_{i-1}}|$, and therefore the number of pairs $S,S'$ of Glauber graph vertices that use these edges is at most
$$\sum_{T,T': |\classt{}{T}| \leq |\classt{}{T_{i-1}}|}|\classt{}{T}||\classt{}{T'}| \leq N|\mathcal{S}||\classt{}{T_{i-1}}| = O(1)N|\classt{}{T_{i-1}}|.$$

Therefore, since there are $|\classt{}{T_{i-1}}|$ edges between $\classt{}{T_{i-1}}$ and $\classt{}{T_i}$ (by Condition~\ref{maincond4}), each such edge carries at most $N|\mathcal{S}||\classt{}{T_{i-1}}|/|\classt{}{T_{i-1}}| = O(N)$ units of flow, giving $O(1)$ congestion.

To bound congestion within $\classt{}{T_i}$, we specify the routing of flow from $\mathcal{Y}_i$ (the set of vertices on the~$\classt{}{T_{i-1}}, \classt{}{T_i}$ boundary) to $\mathcal{Z}_i$ (the set of vertices on the $\classt{}{T_i}, \classt{}{T_{i+1}}$ boundary) as follows: first let each $Y \in \mathcal{Y}_i$ send an equal fraction of its flow\textemdash of which it receives $O(N)$ units from each of $O(1)$ edges\textemdash to \emph{every vertex in $\classt{}{T_i}$}, using $\rho_{T_i}$ to route the flow. Then let each $Z \in \mathcal{Z}_i$ receive its flow similarly from all vertices in~$\classt{}{T_i}$. The resulting congestion across each edge is at most 
$$(2O(N)/|\classt{}{T_i}|)\cdot c^{\log n - 1}|\classt{}{T_i}|/N \leq c^{\log n},$$
for a constant $c$. This gives the desired congestion bound, proving the lemma.
\end{proof}

We now prove Lemma~\ref{lem:hierexact} by tracing the polynomial factors in the proof of Lemma~\ref{lem:altcondexp}:
\lemhierexact*
\begin{proof}
The analysis is similar to the proof of Lemma~\ref{lem:nonhierexact}, with the following modifications: each edge set $\edget{}{T}{T'}$ from $\classt{}{T}$ to a parent $\classt{}{T'}$ has $|\edget{}{T}{T'}| = |\classt{}{T}|$. Therefore, outbound flow along each edge in such an edge set is at most $N|\classt{}{T}|/|\edget{}{T}{T'}| = N$: each vertex (all vertices in~$\classt{}{T}$ are boundary vertices) then receives from each other vertex at most $N/|\classt{}{T}|$ units. As we will show shortly (see analysis of through flow below), edges to children each carry at most $K|\classt{}{T}|$. Thus we will count the flow resulting from edges to children with through flow.

Inbound flow is symmetric. The result is to scale the amount of flow across each edge internal to $\classt{}{T}$ by a factor of $2N/|\classt{}{T}|$.

For through flow (including the outbound flow to children as described above), each boundary vertex in $\classt{}{T}$ carrying flow from (or to) a set of child classes $\{\classt{}{T_1'}, \dots, \classt{}{T_k'}\}$ carries at most $\sum_{i=1}^kN|\classt{}{T_i'}/(|\classt{}{T}|) \leq NK_i|\classt{}{T}|/|\classt{}{T}|$ units, where $K_i < K$ is the number of classes descendent from $\classt{}{T_i'}$, including $\classt{}{T_i'}$ itself. This sum is at most $N(K-1)$. Each boundary vertex carrying flow from (or to) an ancestor similarly carries at most $NK$ units. Thus through flow contributes a factor of $2N(K-1)/|\classt{}{T}|$.

The resulting overall congestion is therefore at most
$$O((2K+1)^{2\log n}),$$ where the extra~$+1$ term comes from accounting for the congestion from routing flow for pairs in the same class, and applying Lemma~\ref{lem:flowexp} and Lemma~\ref{lem:expmixing} gives the resulting mixing bound.
\end{proof}

\label{sec:mainfwbounds}
\subsection{Independent sets}
\label{sec:indsetexp}
We now finish the proof of the unbiased case of Theorem~\ref{thm:indsetmix}.

\subsubsection{Verification of conditions}
To show that the chain on independent sets satisfies the conditions of the hierarchical framework when treewidth is bounded (but degree is unbounded), we first define a partial order $\prec$ on the classes in $\classesx{IS}{G}$. Recall (Definition~\ref{def:ispart}) that these are the classes induced by the separator $X$ in the underlying graph $G$.
\begin{definition}
\label{def:indsetpo}
For $\classt{IS}{T}, \classt{IS}{T'} \in \classesx{IS}{G}$, let $\classt{IS}{T} \prec \classt{IS}{T'}$ if $T \subseteq T'$ and $T \neq T'$. Call~$\classt{IS}{T}$ an \emph{ancestor} of $\classt{IS}{T'}$, and $\classt{IS}{T'}$ a \emph{descendant} of $\classt{IS}{T}$. If $\classt{IS}{T}$ covers $\classt{IS}{T'}$ in this relation, call $\classt{IS}{T}$ a \emph{parent} of $\classt{IS}{T'}$, and $\classt{IS}{T'}$ a \emph{child} of $\classt{IS}{T}$.
\end{definition}

We now prove that the chain on independent sets satisfies the conditions of the hierarchical framework on graphs of bounded treewidth.

\begin{lemma}
\label{lem:indsetmaincond}
Given a graph $G$ with fixed treewidth $t - 1$, the hardcore Glauber dynamics on the independent sets of $G$ satisfies the conditions of the hierarchical framework.
\end{lemma}
\begin{proof}
Let $\flipg{IS}{G}$, $X$, and $\classesx{IS}{G}$ be as previously defined. We have already proven Condition~\ref{maincond1} and Conditions~\ref{maincond4} through~\ref{maincond7} in Lemmas~\ref{lem:iscondnumclasses} through~\ref{lem:iscondsesizes}.

The partial order in Definition~\ref{def:indsetpo} satisfies Condition~\ref{maincond2} because for every parent class $\classt{IS}{T}$ and child class $\classt{IS}{T'}$, the recursive subproblems in the Cartesian product comprising $\classt{IS}{T'}$ are at least as constrained as the subproblems in the product comprising $\classt{IS}{T}$. That is, $\classt{IS}{T}$ and~$\classt{IS}{T'}$ are each a Cartesian product of two smaller Glauber graphs on the independent sets in subgraphs~$A_T$ and~$B_T$ of~$G$, and subgraphs~$A_{T'}$ and~$B_{T'}$ of~$G$ respectively. We have $V(A_{T'}) \subseteq V(A_T)$ and $V(B_{T'}) \subseteq V(B_T)$, where the set $V(A_T) \setminus V(A_{T'})$ consists of the vertices in~$A$ that have a neighbor in~$T'$ but not in~$T$.

Condition~\ref{maincond3} follows from the fact that the empty independent set is the unique set that is an ancestor of all other independent sets.

\end{proof}
It now follows by Lemma~\ref{lem:flowexp} that $\flipg{}{G}$ has expansion $\Omega(1/n^{O(1)})$, and Theorem~\ref{thm:indsetmix} follows from this fact and from Lemma~\ref{lem:expmixing}. More precisely, observing that the number of classes in the partition is at most $2^{t+1}$ and applying Lemma~\ref{lem:hierexact} gives the bound claimed in Theorem~\ref{thm:indsetmix}, namely
$$O(((1+\hat\lambda)\hat\lambda)^2(1+\log \hat\lambda)n^{2(t+2)(1+\log \hat\lambda)+5}),$$
where~$\hat\lambda = \max\{\lambda, 1/\lambda\}.$

\subsection{Partial $q$-colorings}
\label{sec:partialqhier}
We now prove the unbiased case of the claim about partial colorings in Theorem~\ref{thm:lcolor}:

\begin{definition}
Let $\flipg{PCOL}{G,L}$, given an input graph $G$ and function $L: V(G) \rightarrow 2^{[q]}$, be the graph whose vertices are the partial list colorings of $G$, and whose edges are the pairs of partial list colorings that differ by the removal or addition of a color assignment to a single vertex.
\end{definition}

We show that this Glauber graph satisfies the conditions of the hierarchical framework:
\begin{lemma}
Given a graph $G$ with bounded treewidth and degree and list function $L: V(G) \rightarrow 2^{[q]}$, where $q \geq \Delta + 2$ is fixed and $L(v) \geq \delta(v) + 2$ for all $v \in V(G)$, the Glauber graph $\flipg{PCOL}{G,L}$ has expansion $\Omega(1/n^c)$, where $c = O(1)$.
\end{lemma}
\begin{proof}
The partitioning is the same as in the proof of Lemma~\ref{lem:colexp}, except that we allow each class to be identified with a partial list coloring of $X$. Condition~\ref{maincond1}, Condition~\ref{maincond5}, Condition~\ref{maincond6}, and Condition~\ref{maincond7} are as before. For Conditions~\ref{maincond2} and~\ref{maincond3}, the partial order is analogous to the partial order for independent sets: given partial list colorings $C$ and $C'$ of $X$, let $C$ be a parent of $C'$ if $C$ and $C'$ agree except for a single vertex to which $C'$ assigns a coloring and $C$ does not. Condition~\ref{maincond4} follows from this definition. The lemma follows.
\end{proof}

We obtain the bound in Theorem~\ref{thm:lcolor} via the observations in Section~\ref{sec:mixingder}.

\section{All $\lambda > 0$}
\label{sec:alllambda}
Until now, we have only considered the unbiased versions of our chains. In this section we complete the proof of Theorem~\ref{thm:indsetmix}, for arbitrary fixed $\lambda > 0$. To do so, we need to introduce the standard notion of \emph{conductance} \cite{sinclair_1992}, which extends the definition of expansion in the natural way to the setting of a weighted graph.

\subsection{(Weighted) Conductance}
\label{sec:prelimrev}
The conductance is defined with respect to a \emph{stationary} distribution $\pi$ induced by a random walk. The stationary distribution is the distribution to which the random walk converges in the limit. The convergence requires mild conditions: (i) that walk be \emph{ergodic}, meaning that the Glauber graph is connected; (ii) that the walk be \emph{reversible}; and (iii) that the walk be \emph{lazy}.

Laziness means that with constant probability the walk stays at the current vertex at any step; reversibility means that for every pair of sets $S, S' \in \flipg{IS}{G}$, we have 
$$\pi(S)P(S, S') = \pi(S')P(S', S),$$
where $P(S, S')$ denotes the probability that $X_{t+1} = S'$, given that $X_t = S$.

The Glauber dynamics on independent sets is known to satisfy these conditions, and it is easy to show that our other Glauber dynamics satisfy them as well.

In the case of the Glauber dynamics on independent sets, the stationary distribution $\pi$ evaluates to
$$\pi(S) = \lambda^{|S|}/Z(\flipg{IS}{G}),$$ 
where for each of our Glauber graphs $\flipg{}{G}$, $$Z(\flipg{}{G}) = \sum_{S\subseteq V(\flipg{}{G})} \lambda^{|S|}$$ is the normalizing constant.
For all independent sets $S$ in $G$, and for all $S'$ such that $|S \setminus S'| = 1,$ we have
$$\pi(S)P(S, S') = \pi(S')P(S', S) = ( \frac{1}{n(1+\lambda)})(\frac{\lambda^{|S|}}{Z(\flipg{IS}{G})}),$$
where $n = |V(G)|$.

For dominating sets and partial $q$-colorings, we define the same distribution; for $b$-edge covers we define the analogous distribution over edges.

\begin{remark}
\label{rmk:weights}
For each of our Glauber graphs $\flipg{}{G}$, the probability transition function $P(S, S')$, viewed as a matrix, is in fact the adjacency matrix of an edge-weighted version of $\flipg{}{G}$, ignoring self loops.
\end{remark}

That is:
\begin{definition}
\label{def:weightedflipg}
Given a Glauber graph $\flipg{}{G}$ and a Markov chain on $\flipg{}{G}$ with stationary distribution $\pi$ and probability transition function $P$, assign the weight $\pi(S)$ to each vertex~$S$ of~$\flipg{}{G}$, and assign the weight $Q(S, S') = \pi(S)P(S, S')$ to each edge $(S, S')$.
\end{definition}

\begin{definition}
\label{def:weightedcart}
Extend the definition of a Cartesian graph product to the weighted graphs described in this section, so that for vertices $g \in V(G), h \in V(H)$, the weight of the tuple $(g, h) \in V(G \Box H)$ is $\pi(g, h) = \pi_G(g)\pi_H(h)$, where $\pi_G$ and $\pi_H$ are the vertex weight functions for $G$ and $H$ respectively. Let the weight of each edge $e$ between $(g, h)$ and $(g', h')$ be
$$Q(e) = \pi_H(h)(\Delta_G Q_G(g, g'))/(\Delta_G + \Delta_H),$$ if $g \neq g'$ and $h = h'$, and
$$Q(e) = \pi_G(g)(\Delta_H Q_H(h, h'))/(\Delta_G + \Delta_H),$$ if $g = g'$ and $h \neq h'$, where $Q_G$ and $Q_H$ are the edge weight functions for~$G$ and~$H$, and~$\Delta_G$ and~$\Delta_H$ are the maximum degrees of $G$ and $H$.

For the self loop $e = ((g, h), (g, h)),$ let

$$Q(e) = \pi(g, h) - \sum_{(g'', h): g'' \neq g} Q((g, h), (g'', h)) - \sum_{(g, h''): h'' \neq h} Q((g, h), (g, h'')),$$

\end{definition}

\begin{lemma}
\label{lem:weightfactor}
Given Definition~\ref{def:weightedcart}, the stationary distribution $\pi$ in the discussion leading to Remark~\ref{rmk:weights}, and the resulting vertex and edge weights as in Definition~\ref{def:weightedflipg}, for each of our Glauber graphs $\flipg{}{G}$ and for each class $\classt{}{T} \cong \flipg{}{A} \Box \flipg{}{B}$, and for each $S \in V(\flipg{}{A}), S' \in V(\flipg{}{B})$, the following facts hold:
\begin{enumerate}
\item The vertex weight of $S \cup S' \cup T$ in $\flipg{}{G}$ is equal to 
$$\pi_{\flipg{}{G}}(S \cup S' \cup T) = \pi_{\classt{}{T}}(S, S')\pi_{\flipg{}{G}}(\classt{}{T}),$$ where $\pi_{\flipg{}{G}}(\classt{}{T})$ is defined as $\sum_{U\in \classt{}{T}} \pi_{\flipg{}{G}}(U)$, and
\item For all $S''$ with $|S \setminus S''| = 1$, the weight in $\flipg{}{G}$ of the edge $e$ between $S \cup S' \cup T$ and $S'' \cup S' \cup T$ is
$$Q_{\flipg{}{G}}(e) = Q_{\classt{}{T}}(e)\pi_{\flipg{}{G}}(\classt{}{T})\frac{|V(A)|+|V(B)|}{|V(G)|}.$$
\end{enumerate}
\end{lemma}
\begin{proof}
We have $$\frac{\lambda^{|S|+|S'|+|T|}}{Z(\flipg{}{G})} = \pi_{\flipg{}{A}}(S)\pi_{\flipg{}{B}}(S')\lambda^{|T|}\cdot \frac{Z(\flipg{}{A})Z(\flipg{}{B})}{Z(\flipg{}{G})} = \pi_{\classt{}{T}}(S, S')\pi_{\flipg{}{G}}(\classt{}{T}),$$ and 
$$Q(e) = \frac{1}{|V(G)|(\lambda+1)}\cdot \frac{\lambda^{|S|+|S'|+|T|}}{Z(\flipg{}{G})} = Q_{\flipg{}{A}}(S, S'')\pi_{\flipg{}{B}}(S')\lambda^{|T|}\cdot \frac{Z(\flipg{}{A})Z(\flipg{}{B})}{Z(\flipg{}{G})}\cdot \frac{|V(A)|}{|V(G)|}$$
$$= Q_{\classt{}{T}}(e)\lambda^{|T|}\cdot \frac{Z(\flipg{}{A})Z(\flipg{}{B})}{Z(\flipg{}{G})}\cdot \frac{|V(A)| + |V(B)|}{|V(G)|}$$
$$= Q_{\classt{}{T}}(e)\pi_{\flipg{}{G}}(\classt{}{T})\frac{|V(A)|+|V(B)|}{|V(G)|}.$$
\end{proof}

\begin{definition}
\label{def:conductance}
Given a lazy, reversible, ergodic random walk on a weighted graph $\mathcal{M} = (\mathcal{V}, \mathcal{E})$ with stationary distribution $\pi$ and probability matrix $P:\mathcal{V} \times \mathcal{V} \rightarrow [0, 1]$, the \emph{conductance} is the quantity $$\phi(\mathcal{M}) = \min_{\mathcal{S} \subseteq \mathcal{V}: 0 < \pi(\mathcal{S}) \leq 1/2} \frac{Q(\mathcal{S}, \mathcal{V} \setminus \mathcal{S})}{\pi(\mathcal{S})},$$
where for sets $\mathcal{S}, \mathcal{S}' \subseteq \mathcal{V}$,
$$\pi(\mathcal{S}) = \sum_{S\in\mathcal{S}}\pi(S),$$
and $$Q(\mathcal{S}, \mathcal{S}') = \sum_{S\in \mathcal{S}, S' \in \mathcal{S}'} Q(S, S'),$$
and where $Q(S, S') = \pi(S)P(S, S')$ given $S, S' \in \mathcal{V}$.
\end{definition}

We now extend the definitions of multicommodity flows and congestion:
\begin{definition}
\label{def:mcflowweighted}
Let a multicommodity flow $\rho$ in a graph $\mathcal{M} = (\mathcal{V}, \mathcal{E})$ be defined as before, except that each pair of vertices $S, S' \in \mathcal{V}$ exchanges $\pi(S)\pi(S')$ units of flow in each direction. Let 
$$\rho_{max} = \sum_{e\in\mathcal{E}} \frac{\rho(e)}{Q(e)}.$$
\end{definition}

The following generalizations of Lemma~\ref{lem:flowexp} and Lemma~\ref{lem:expmixing} relate the conductance, congestion, and mixing time~\cite{sinclair_1992}:

\begin{theorem}
\label{thm:mcfcond}
Given a multicommodity flow~$\rho$ in a graph~$\mathcal{M}$, the conductance~$\phi$ satisfies~$\phi \geq 1/(2\rho)$.
\end{theorem}

\begin{theorem}
\label{thm:weightedmix}
The mixing time of a Markov chain with state space $\Omega$, stationary distribution $\pi^*$, and conductance at least $\phi$ is at most
$$\tau(\varepsilon) = O(\phi^{-2}\log (1/(\pi_{\min}^*\cdot\varepsilon))),$$
where $$\pi_{\min}^* = \min_{u \in \Omega} \pi^*(u).$$
\end{theorem}

\subsection{Analysis of flow construction}
We now complete the proof of Theorem~\ref{thm:indsetmix}. It suffices to show the following lemma:

\begin{lemma}
\label{lem:alllambdais}
The flow $\rho$ constructed in $\flipg{IS}{G}$ in the proof of Lemma~\ref{lem:altcondexp}, adjusted so that $\flipg{IS}{G}$ is weighted according to the parameter $\lambda > 0$, and so that each pair of sets $S, S'$ exchanges $\pi(S)\pi(S')$ units of flow as in Definition~\ref{def:mcflowweighted}, results in a congestion factor gain of at most $\rho_{\max} = O(1)$ at each of the $O(\log n)$ levels of induction, resulting in at most polynomial overall congestion. The same holds for the chain on partial $q$-colorings.
\end{lemma}
\begin{proof}
We use the same inductive argument, with the following adjustment: if  $\classt{}{T}$ is a descendant of $\classt{}{T_{i-1}}$, and $\classt{}{T_{i-1}}$ is a child of $\classt{}{T_{i}}$, where $\classt{}{T}$ uses the edges between $\classt{}{T_{i-1}}$ and $\classt{}{T_{i}}$ to send flow to $\classt{}{T'}$, then distribute this flow as before across these edges, but now let each edge carry flow in proportion to its weight. We have $\pi(\classt{}{T}) = O(1)\pi(\classt{}{T_{i-1}})$\textemdash because for every independent set $S \in \classt{}{T}$ there exists a distinct independent set $S' \in \classt{}{T_{i-1}}$ with $\pi(S') = (1/\lambda^{|T\setminus T_{i-1}|})\pi(S)$, namely $S' = S \setminus (T \setminus T_{i-1})$.

Each edge $(S, S')$ with $S \in \classt{}{T_{i}}, S' \in \classt{}{T_{i-1}}$, satisfies $Q(S, S') = \Omega(\pi(S)/n)$ (where the constant-factor difference depends on $\lambda$). Thus the congestion along these edges is still $O(n)$. We then allow each vertex $S \in \classt{}{T_{i}}$, having received at most $O(\pi(S))$ units along each of $O(1)$ incoming edges from child classes, to distribute these units to all other vertices in $\classt{}{T_{i}}$ according to their weight. That is, let $S$ send $O(\pi(S)\pi(S'')/\pi(\classt{}{T_{i}}))$ units to each $S'' \in \classt{}{T_{i}}$. By the inductive hypothesis and Lemma~\ref{lem:cartflow}, a flow $\rho_{T_{i}}$ already exists under which $S$ sends $\pi(S)\pi(S'')/(\pi(\classt{}{T_{i}}))^2$ units to $S''$ at a congestion cost of $O(nc^{\log n - 1})$, for appropriate constant $c$. Thus letting $S$ send $O(\pi(S)\pi(S'')/\pi(\classt{}{T_{i}}))$ units to $S''$ reduces the amount sent across each edge by at least a factor of~$1/\pi(\classt{}{T_{i}})$, while the weight of each edge increases when passing from the factor graphs of~$\classt{}{T_{i}}$ to~$\flipg{}{G}$ by at most the same factor\textemdash up to the change in degree of the Glauber graph\textemdash by Lemma~\ref{lem:weightfactor}. This gives congestion cost at most $O(nc^{\log n})$.
\end{proof}

\subsubsection{Specific polynomial bounds}
We now revisit the discussion in Section~\ref{sec:mixingder}. In Theorem~\ref{thm:indsetmix}, the $1 + \log \hat\lambda$ term in the exponent comes from observing that, in Lemma~\ref{lem:hierexact}, we can replace the~$K + 1$ term with $K\lambda^{t+1} + 1$ \textemdash since in the proof of Lemma~\ref{lem:hierexact}, this is the factor by which the flow carried into a class from a child class increases when adjusting for the weights induced by the parameter~$\lambda$. A similar analysis gives the result for partial $q$-colorings in Theorem~\ref{thm:lcolor}.

\subsubsection{The Ising and Potts models}
\label{sec:isingpotts}
As we discussed in the introduction, one can apply our framework to the Ising and Potts models when the parameters of these models are fixed. We do not give the standard definitions of these models or a detailed proof. Instead, we simply observe that in a graph of bounded treewidth and degree, the same decomposition technique given in the non-hierarchical framework applies, with the following modification. Since all assignments of spins are possible, so instead of considering the cardinalities of the sets we consider weights of configurations under the standard (exponentiated) energy functions. It is easy to verify the conditions under this modification, using the insight that the weights of the classes differ by a constant factor from one another, since this factor is determined only by evaluating the energy function at a constant number of edges and vertices.

\section{Dealing with non-independence}
\label{sec:adapting}
\label{sec:nonind}
The Glauber dynamics on independent sets induces a Glauber graph, $\flipg{IS}{G}$, that behaves well when partitioned into classes. That is, each class $\classt{IS}{T}$ is isomorphic to the Cartesian product of two Glauber graphs on subgraphs of $G$. As we will see in Appendix~\ref{sec:mainfwcond}, the Glauber dynamics on partial $q$-colorings is similarly well-behaved. Unfortunately, as we will discuss in Appendix~\ref{sec:adapting}, this does not hold for dominating sets or, in the unbounded-degree case, for $b$-edge covers. In these problems, the selection $T \subseteq X$ of vertices (or edges) in the separator $X$ with which the class $\classt{}{T}$ is identified imposes constraints on what vertices (or edges) can be chosen in the two subgraphs $A \cup B = G \setminus X$\textemdash and choices in $A$ may invalidate those in $B$.

In the case of maximal independent sets and maximal $b$-matchings, the situation is worse: the classes induced by selection of $T \subseteq X$ may not even be internally connected.

We address both of these problems by relaxing the framework condition that each class $\classt{}{T}$ be a Cartesian product of Glauber graphs, and instead require that each class $\classt{}{T}$ be the (not necessarily disjoint) union of Cartesian products of Glauber graphs, satisfying certain conditions. We fully specify this condition, and show that the remaining chains satisfy it, this section. That is, we complete the proofs of Theorem~\ref{thm:domsetmix} and Theorem~\ref{thm:mindsetmix}. The principal problem is that when attempting to partition the Glauber graph into classes as we did for independent sets, the resulting classes are not isomorphic to Cartesian products of Glauber graphs. For instance, in the case of $b$-edge covers, we wish to identify a class of $b$-edge covers with the set $T$ of edges selected within the subgraph of $G$ induced by the separator $X$. Unfortunately, the resulting subproblems on $A$ and $B$ (where, as before, $A \cup B = V(G) \setminus X$) are not independent. This is because for each vertex $x \in X$, the sum of the number of incident edges selected in $A$ and those in $B$ must be at least $b(x)$, so the choices made in $A$ depend on those made in $B$, and vice versa.

The solution is as follows: we divide each class into smaller (not necessarily disjoint) ``subclasses'', each of which is a Cartesian product of smaller Glauber graphs on $b$-edge covers. We detail this in Appendix~\ref{sec:becfull}.

We encounter a similar problem in the case of dominating sets, with an additional challenge that will require us to generalize the definition of a dominating set into what we call the ``constrained Steiner dominating set'' problem. We give the full details in Appendix~\ref{sec:csds}.

For maximal independent sets and maximal $b$-matchings (Appendix~\ref{sec:mismaxandb}), the non-hierarchical framework is more natural, as we require bounded degree. The challenge is twofold: first, we need to define the Glauber graphs and show that they are connected. Secondly, we need to deal with non-independence as with $b$-edge covers and dominating sets\textemdash with the additional challenge, as we will see, that the classes are not necessarily internally connected.

\subsection{Framework relaxation to allow non-independence}
\label{sec:nonindcond}
\begin{lemma}
\label{lem:nonindhiercond}
Suppose a Glauber graph $\flipg{}{G}$ satisfies the conditions of the hierarchical framework in Appendix~\ref{sec:mainfwcondlist}, except for Condition~\ref{maincond6}. Suppose further that each class $\classt{}{T} \in \mathcal{S}$ is the union of at most $O(1)$ subclasses $\classt{}{T} = \classt{}{T_1} \cup \classt{}{T_2} \cup \cdots \classt{}{T_k}$, where for $i = 1, \dots, k - 1$:
\begin{enumerate}
\item $|\classt{}{T_i}| = \Theta(1)|\classt{}{T_{i+1}}|$, and
\item $\classt{}{T_i}$ and $\classt{}{T_{i+1}}$ share at least $\Omega(1)|\classt{}{T_i}| = \Omega(1)|\classt{}{T_{i+1}}|$ vertices.
\end{enumerate}
Suppose further that for $i = 1, \dots, k$, $\classt{}{T_i}$ is isomorphic to the Cartesian product of two Glauber graphs $\flipg{}{G_1}$ and $\flipg{}{G_2}$, each of which can be recursively partitioned in the same way as $\flipg{}{G}$.

Then the expansion of $\flipg{}{G}$ is $\Omega(1/n^c)$, where $c = O(1)$.
\end{lemma}
\begin{proof}
It suffices to construct a multicommodity flow among the subclasses in $\classesx{}{T}$ and bound its congestion. By the inductive hypothesis and Lemma~\ref{lem:cartflow}, each subclass $\classt{}{T_i}$ has an internal flow $\rho_{T_i}$ with congestion $\rho_{T_i\max} \leq c^{\log n-1}$. We would like to derive a flow $\rho_T$ with congestion $\rho_{T\max} \leq c^{\log n-1}$; this will allow the rest of the proof of Lemma~\ref{lem:altcondexp} to be applied.

The solution is to follow the proof sketch of Lemma~\ref{lem:iscondexp}: for Glauber graph vertices $S \in \classt{}{T_i}, S' \in \classt{}{T_j} \neq \classt{}{T_i}$, send the $S-S'$ flow through the classes $\classt{}{T_{i+1}}, \classt{}{T_{i+1}}, \dots, \classt{}{T_{j-1}}$. For $l = i, \dots, j$, let $\mathcal{Y}_l$ and $\mathcal{Z}_l$ be as in the proof of Lemma~\ref{lem:iscondexp}, except that $Z_l = Y_{l+1}$. That is, the boundary vertices in consecutive pairs of classes on the path are shared between the two classes. The routing of flow within each class on the path is the same as in Lemma~\ref{lem:iscondexp}.

The resulting congestion bound is the same as in Lemma~\ref{lem:iscondexp}. The only concern is that since the subclasses may not be disjoint, each edge within a subclass may incur congestion from multiple steps on the path. However, because the number of classes is $O(1)$, there are $O(1)$ such steps, and thus the factor by which this increases congestion is $O(1)$.
\end{proof}

\subsection{$b$-edge covers in the relaxed hierarchical framework}
\label{sec:becfull}
To finish the proof of Theorem~\ref{thm:domsetmix}, it now suffices to show that the chains on dominating sets and $b$-edge covers satisfy the conditions of the hierarchical framework when treewidth is bounded. We begin with $b$-edge covers.

Let $\classt{BEC}{T}$ be defined as in Section~\ref{sec:bprobs}, with the following modification: define each class $\classt{BEC}{T}$ so that $T$ is identified with the set of selected edges \emph{both of whose endpoints} are in $X$, instead of including all edges incident to vertices in $X$.

We now divide each class $\classt{BEC}{T}$ into subclasses. For each $x \in X$, let $\delta_T(x)$ be the number of edges incident to $x$ (from neighbors in $X$) that are selected in $T$. Let $b'(x) = b(x) - \delta_T(x)$, i.e. decrease~$b(x)$ by the number of edges incident to $x$ selected in $X$. For each $x \in X$, every valid $b$-edge cover in $T$ includes numbers of edges from neighbors in $A$ and $B$ that sum to at least $b'(x)$.

We will define a subclass of $\classt{BEC}{T}$ for each possible assignment of $b$-values to the vertices in $X$ in the subproblems on $A$ and $B$. (The number of these subclasses, since $|X| = O(1)$ and $b$ is bounded, is still $O(1)$.)

Formally: 
\begin{definition}
\label{def:beta}
Define functions $\beta$ and $\overline{\beta}$ as any assignments of $b$-values, in the subproblems on~$A$ and~$B$ respectively, to all vertices $x \in X$, such that the $\beta$ and $\overline{\beta}$ values sum to~$b'(x)$ for each~$x$.
\end{definition}

There are many degrees of freedom in defining $\beta$. Consider each possible choice of $\beta$ and $\overline{\beta}$.

\begin{definition}
\label{def:tbeta}
Define the subclass $\classt{BEC}{T_\beta}$ as the set of all $b$-edge covers that consist of a $\beta$-edge cover in $A$ and a $\overline{\beta}$-edge cover in $B$.
\end{definition}

That is, in class $\classt{BEC}{T_\beta}$, for each~$x$, the number of incident edges selected in $A$ is at least $\beta(x)$, and the number of incident edges in $B$ is at least $\overline{\beta}(x)$.

Each of these subclasses $\classt{BEC}{T_\beta}$ is a Cartesian product of $b$-edge cover Glauber graphs, over subgraphs $A$ and $B$ of $G$, and thus internally has a good flow $\rho_\beta$; thus it suffices to combine flows within these subclasses together to design a flow $\rho_T$ in $\classt{BEC}{T}$. We can then apply the hierarchical framework to obtain the desired flow in $\flipg{BEC}{G}$.

\begin{lemma}
\label{lem:bedgesubflow}
Given a graph $G$ and corresponding Glauber graph $\flipg{BEC}{G}$, each class $\classt{BEC}{T}$ of $b$-edge covers in $\flipg{BEC}{G}$ satisfies the conditions of Lemma~\ref{lem:nonindhiercond}.
\end{lemma}
\begin{proof}
The number of subclasses is~$O(1)$. The subclasses are also all within an $O(1)$ size factor of one another. To see this, compare $|\classt{BEC}{T}|$ and $|\classt{BEC}{T_\beta}|$, for any $\beta$. Fix some lexicographic ordering of the edges of $G$. For every $b$-edge cover $S \in \classt{BEC}{T}$, there exists a $b$-edge cover $S' \in \classt{BEC}{T_\beta}$ that includes the lexicographically first $\beta(x)$ edges in $A$ incident to $x$, for each~$x \in X$, and also includes the first $\overline{\beta}(x)$ edges in $B$ incident to $x$. (Let $S'$ agree with $S$ on all other edges.) This is a $2^b$-to-1 mapping, i.e. an $O(1)$-to-1 mapping.

Finally, every pair of subclasses overlaps in at least $\Omega(1)|\classt{BEC}{T}|$ vertices: consider the set of all $b$-edge covers in $\classt{BEC}{T}$ in which for each $x \in X$, $x$ has $\min\{b'(x), \delta_A(x)\}$ incident edges selected in $A$, and $\min\{b'(x), \delta_B(x)\}$ incident edges selected in $B$. The number of such $b$-edge covers is $\Omega(1)|\classt{BEC}{T}|$, by similar reasoning to the above; furthermore, every pair of subclasses of $\classt{BEC}{T}$ contains this set of $b$-edge covers. The lemma follows.
\end{proof}

The rest of the hierarchical framework conditions are easy to verify, and thus the result in Theorem~\ref{thm:domsetmix} for the unbiased case of $b$-edge covers follows. The specific bound follows from the following observations: first, constructing the flow within a class $\classt{BEC}{T}$ incurs a factor of 
$$2L|\classt{BEC}{T}|/(\nu(\min_\beta|\classt{BEC}{T_\beta}|)),$$
where $\nu$ is the minimum fraction of vertices shared by a pair of adjacent classes whose intersection is used in the flow, $L$ is the maximum number of subclasses that a vertex can belong to, and $\min_\beta|\classt{BEC}{T_\beta}|$ is the smallest subclass of $\classt{BEC}{T}$. $L$ is at most the number of subclasses, which is upper-bounded by $(b+1)^{t+1}$; the smallest subclass has size at least $|\classt{BEC}{T}|/2^{b(t+1)}$; and $\nu \geq 1/2.$ (The latter two facts follow readily from observations made in the proof of Lemma~\ref{lem:bedgesubflow}.)

Thus the construction of the flow within $\classt{BEC}{T}$ incurs a factor of at most $4(b+1)^{t+1}\cdot 2^{b(t+1)}$. In the biased chain the overlap between adjacent subclasses is at least $\lambda/(1+\lambda)$ instead of $1/2$, and we need to adjust the ratio $|\classt{BEC}{T}|/|\classt{BEC}{T_\beta}|$ by a factor of $\hat\lambda^{b(t+1)}.$ Therefore this expression becomes
$$2\frac{1+\lambda}{\lambda}(b+1)^{t+1}\cdot (2\hat\lambda)^{b(t+1)}.$$

The rest of the inductive step is as in the case of independent sets, i.e. we apply Lemma~\ref{lem:hierexact}, using $K \leq 2^{t(t+1)/2},$ $\Delta_{\mathcal{M}} \leq m$, and $N \leq 2^m$. (In the weighted case, $\pi_{\min}^* \leq (2\hat\lambda)^m.$ When considering $K$ in the application of Lemma~\ref{lem:hierexact}, we need to weight $K$ by a factor of $\hat\lambda^{t(t+1)/2}$.

Thus we obtain an additional factor of
$$2(K\hat\lambda^{t(t+1)/2} + 1) \leq 2((2\hat\lambda)^{t(t+1)/2+1}).$$

Altogether, these two flow constructions combined, in each iteration, result in a factor of at most
$$4(\frac{1+\lambda}{\lambda})(b+1)^{t+1}\cdot (2\hat\lambda)^{(t+1)(b+t/2)+1}.$$

The resulting mixing time is therefore at most
$$O(((1+\hat\lambda)\hat\lambda)^2m^3(\log\hat\lambda+1)n^{2(2+\log(1+\lambda)-\log(\lambda)+(t+1)\log(b+1)+((t+1)(b+t/2)+1)(1+\log\hat\lambda))}).$$

We have ignored one detail: technically the number of levels of induction is greater than $\log n$, because the $t+1$ vertices in the separator are included in the independent subproblems within each subclass. Furthermore, we cannot assume that we have two connected components of size at most $|V(G)|/2$ at each level of decomposition, so the base of the log is 3/2 and not 2.

However, for every $\varepsilon < 1/2,$ we have for all $n \geq (t+1)/\varepsilon$ that $2n/3 + t + 1 \leq n(2/3+\varepsilon)$. Thus at the cost of a base case for the induction of $(2\hat\lambda)^{((t+1)/\varepsilon)^2}$, we adjust the $\log n$ exponent in the congestion term to $\log_{1/(2/3+\varepsilon)} n$. Letting $\varepsilon = 1/6$, we obtain the mixing bound claimed in Theorem~\ref{thm:domsetmix}, namely

$$O\left(((2\hat\lambda)^{36(t+1)^2})^2\left(\frac{\hat\lambda}{(1+\hat\lambda)}\right)^2 m^3(\log\hat\lambda+1)n^{\frac{2(3+\log(1+\lambda)-\log(\lambda)+(t+1)\log(b+1)+((t+1)(b+t/2)+1)(1+\log\hat\lambda))}{\log(6/5)}}\right).$$

\subsection{Dominating sets in the relaxed hierarchical framework}
\label{sec:csds}
To finish the proof of Theorem~\ref{thm:domsetmix} in the unbiased case, we now deal with dominating sets.

As with $b$-edge covers, defining classes in the same way as in the case of independent sets does not result in Cartesian products of dominating set Glauber graphs, because it may be that some vertices in $X$ are not dominated by vertices in $T \cap X$; these vertices must then be dominated by vertices in $A$ or in $B$.

Therefore, to preserve the recursive structure of the problem and thus complete the proof of Theorem~\ref{thm:domsetmix}, we define the \emph{constrained Steiner dominating set problem} as a generalization of the dominating set problem, in which there are three types of vertices:
\begin{enumerate}
\item ``normal'' vertices, which must be dominated and may be selected in a dominating set,
\item ``Steiner'' vertices, which need not be dominated and may be selected, and
\item ``forbidden'' vertices, which must be dominated and must not be selected.
\end{enumerate}

Now, we let $X$ be a balanced vertex separator in $G$ as before. We would like to define each class $\classt{DOM}{T}$ by a choice of vertices in $X$. In the resulting subproblem in $A$ (similarly $B$), we then designate each vertex $v \in A \cup B$ having a neighbor selected in $T \subseteq X$ as Steiner. However, there may be vertices in $X$ that are not selected or dominated by any vertex in $T$. To obtain a valid dominating set, some neighbor of each such vertex $w$ must be chosen in either $A$ or $B$. Thus we have non-independent subproblems, which ruins the Cartesian product structure needed for the divide-and-conquer argument. To resolve this non-independence, we divide $\classt{DOM}{T}$ into subclasses as follows:

\begin{definition}
\label{def:domsubclass}
Given a graph $G$, separator $X$, and CSDS Glauber graph $\flipg{DOM}{G}$, and class~$\classt{DOM}{T}$ of CSDS's in $\flipg{DOM}{G}$, let $U$ be the set of undominated vertices in $X$ in class $\classt{DOM}{T}$. For each subset $W \subseteq U$, let the subclass $\classt{DOM}{T_W}$ be the set of all CSDS's that consist of a union of a CSDS on $A \cup W$, and a CSDS on $B \cup U \setminus W$\textemdash in which each $w \in W$ is a forbidden vertex in the $A$ subproblem, and each $\overline{w} \in U \setminus W$ is a forbidden vertex in the $B$ subproblem.
\end{definition}

There are at most $2^t = O(1)$ such subclasses. Technically, as with $b$-edge covers, these are not equivalence classes, as some CSDS solutions may belong to multiple classes. We will address this shortly, but first we specify how to route flow among the subclasses within $\classt{DOM}{T}$. Once we have specified this flow, we can simply apply the flow described in the proof of Lemma~\ref{lem:altcondexp} to route flow among the ``main'' classes.

\begin{lemma}
\label{lem:domsubflow}
Given a graph $G$, corresponding CSDS Glauber graph $\flipg{DOM}{G}$, and a class~$\classt{DOM}{T}$ of CSDS's in $\flipg{DOM}{G}$, the partition into subclasses given in Definition~\ref{def:domsubclass} satisfies the conditions of Lemma~\ref{lem:nonindhiercond}.

\end{lemma}
\begin{proof}
There are $O(1)$ classes. We observe that the subclasses are all within an $O(1)$ size factor of one another. To see this, compare the sizes of $\classt{DOM}{T}$ and $\classt{DOM}{T_W}$. Since $\classt{DOM}{T_W} \subseteq \classt{DOM}{T}$, $|\classt{DOM}{T_W}| \leq \classt{DOM}{T}$. On the other hand, consider the mapping that sends every CSDS $s \in \classt{DOM}{T}$ to a CSDS $s' \in \classt{DOM}{T_W}$ in which at least one neighbor (say, the first lexicographically) of each $w \in W$ is selected in $A$, and in which the first neighbor of each $\overline{w} \in U \setminus W$ is selected in $B$. This mapping is $2^{|U|}$-to-1 $= O(1)$-to-1, and thus the size factor difference is $O(1)$.

Now, since the subclasses are not equivalence classes, many pairs of subclasses overlap. In particular, let $T$ and $U$ be as before, and suppose for some $u \in U$, $W' = W \cup \{u\}$. Then~$\classt{DOM}{T_W}$ and~$\classt{DOM}{T_{W'}}$ overlap at those CSDS's in which some neighbor of $u$ in $A$ is selected, and some neighbor of $u$ in $B$ is selected. For every such pair of subclasses~$\classt{DOM}{T_W}$ and~$\classt{DOM}{T_{W'}}$, at least half of the CSDS's in $\classt{DOM}{T_W}$ and at least half of those in $\classt{DOM}{T_{W'}}$ are part of the overlap.

$\classt{DOM}{T}$ is internally connected via these overlaps: every CSDS in $\classt{DOM}{T}$ has a path to the trivial CSDS in which every non-forbidden vertex of $A \cup B$ is selected. Thus the conditions of Lemma~\ref{lem:nonindhiercond} are satisfied.
\end{proof}

As in the discussion following Lemma~\ref{lem:bedgesubflow}, we derive the bound in Theorem~\ref{thm:domsetmix} as follows: the flow within a class incurs a congestion factor of 
$$2\frac{1+\lambda}{\lambda}\cdot (2\hat\lambda)^{t+1}|\classt{DOM}{T}|/(\min_W |\classt{DOM}{T_W}|) \leq 2\frac{1+\lambda}{\lambda}(4\hat\lambda)^{t+1}.$$ The application of Lemma~\ref{lem:hierexact} contributes a $2(K\hat\lambda^{t+1}+1)$ factor to the inductive step, with $K = 2^{t+1}$.

Thus the factor for one iteration is at most
$$4\frac{1+\lambda}{\lambda}(8\hat\lambda ^2)^{t+2}.$$

The inclusion of the $t+1$ separator vertices in the subproblems, as with $b$-edge covers, increases the induction depth, and an analogous analysis gives a base case of $(2\hat\lambda)^{6(t+1)}$ using $\varepsilon = 1/6$. Putting all this together with the fact that $\Delta_{\mathcal{M}} \leq n$ and $1/\pi_{\min}^*\leq (2\hat\lambda)^n$ gives a mixing bound of

$$O(((2\hat\lambda)^{6(t+1)})^2(\frac{\hat\lambda}{1+\hat\lambda})^2(1+\log\hat\lambda)\cdot n^{2\log(4\frac{1+\lambda}{\lambda}(8\hat\lambda ^2)^{t+2})/\log(6/5) + 3})$$
$$= O(((2\hat\lambda)^{12(t+1)})(\frac{\hat\lambda}{1+\hat\lambda})^2(1+\log\hat\lambda)\cdot n^{2(2+\log(\frac{1+\lambda}{\lambda}) + (t+2)(3+\log\hat\lambda))/\log(6/5) + 3}).$$

\subsection{Rapid mixing in the relaxed hierarchical framework for all $\lambda > 0$}
\label{sec:relaxedlambda}
We now generalize Lemma~\ref{lem:nonindhiercond} to all $\lambda > 0$, finishing the proof of Theorems~\ref{thm:lcolor} and~\ref{thm:domsetmix}.
\begin{lemma}
For the Glauber graphs $\flipg{BEC}{G}$ and $\flipg{DOM}{G}$, with classes defined as in Lemma~\ref{lem:nonindhiercond}, and with stationary distribution $\pi$ induced by parameter $\lambda$ as in the discussion in Section~\ref{sec:prelimrev}, the flow construction in Lemma~\ref{lem:nonindhiercond} results in a congestion factor gain of at most $\rho_{\max} = O(1)$ at each of the $O(\log n)$ levels of induction, resulting in at most polynomial overall congestion.
\end{lemma}
\begin{proof}
We need to show that the flow construction within a class $\classt{}{T}$ in Lemma~\ref{lem:nonindhiercond} produces at most an $O(1)$-factor increase in congestion; the rest of the argument is similar to the proof of Lemma~\ref{lem:alllambdais}. For the case of dominating sets, consider a pair of $\classt{DOM}{T_W} \subseteq \classt{DOM}{T}$ and $\classt{DOM}{T_{W'}} \subseteq \classt{DOM}{T}$. For every such pair, consider the intersection $\mathcal{I}$ of the two subclasses, namely the set of dominating sets in which for every input graph vertex $v \in W \cup W'$, some neighbor of $v$ is selected in $A$, and for every vertex $w \in U \setminus (W \cap W')$, some neighbor of~$w$ is selected in~$B$. There exists an $O(1)$-to-1 mapping from $\classt{DOM}{T_{W}}$ to $\mathcal{I}$\textemdash found by adding $O(1)$ neighbors of vertices in $U$ as described above to each dominating set $S \in \classt{DOM}{T_W}$\textemdash under which the image~$S'$ of~$S$ has $|S' \setminus S| = O(1)$, and therefore $\pi(S') = \lambda^{|S' \setminus S|}\pi(S) = \Theta(1)\pi(S)$. This shows that $\pi(\mathcal{I}) = \Theta(\pi(\classt{DOM}{T_W})) = \Theta(\pi(\classt{DOM}{T_{W'}}))$.

Thus we use the overlaps between classes to route flow along a path of classes as in the proof of Lemma~\ref{lem:nonindhiercond}. As before, at each class in the path, the internal routing produces an $O(1)$ factor increase in the congestion within the class.
The concern, again, is that due to overlap, there may be edges belonging to multiple classes that thus incur congestion multiple times in the routing of the flow; as before, this is not a problem as there are $O(1)$ pairs of classes for which this occurs.

The argument for $\flipg{BEC}{G}$ is similar, with the intersection $\mathcal{I}$ found by selecting sufficiently many edges incident to each vertex $x \in X$ to satisfy membership in both subclasses~$\classt{BEC}{T_\beta}$ and~$\classt{BEC}{T_{\beta'}}$.
\end{proof}

\subsection{Rapid mixing of the Glauber dynamics on $b$-matchings for all $\lambda > 0$}
For the claim about $b$-matchings in Theorem~\ref{thm:domsetmix}, we do not need the relaxed framework; in fact it suffices to combine Lemma~\ref{lem:bproblemssimple} with the following lemma:
\begin{lemma}
It is easy to adapt the proof of Lemma~\ref{lem:alllambdais} to the hierarchical framework, proving the claim about the Glauber dynamics on $b$-matchings in Theorem~\ref{thm:domsetmix} for all $\lambda > 0$.
\end{lemma}
\begin{proof}
The proof of Lemma~\ref{lem:alllambdais} uses a simple mapping argument to show that for every ancestor~$\classt{}{T_i}$ of a class~$\classt{}{T}$, $\pi(\classt{}{T_i}) = \Theta(1)\pi(\classt{}{T})$, then allows each boundary edge $e$ between classes to carry~$O(1)Q(e)$ units of flow across the boundary, by ensuring that each boundary vertex $S$ carries flow in proportion to its weight $\pi(S)$. Since all pairs of classes $\classt{}{T}, \classt{}{T'}$ have a common ancestor in the case of $b$-matchings, we in fact have $\pi(\classt{}{T}) = \Theta(1)\pi(\classt{}{T'})$ for every pair of classes~$\classt{}{T}, \classt{}{T'}$. The bound on flow across the boundary therefore still holds; the argument for bounding congeston factor increase wthin a class is the same as in the proof of Lemma~\ref{lem:alllambdais}.
\end{proof}
For the specific mixing upper bound for $b$-matchings, we use Lemma~\ref{lem:hierexact}, and observe that $K \leq 2^{\Delta(t+1)}$; the parameter $\lambda$ contributes at most a $\hat\lambda^{\Delta(t+1)}$ factor, $1/\pi_{\min}^* \leq (2\hat\lambda)^m$, and $\Delta_{\mathcal{M}} \leq m$, so we have a mixing upper bound of
$$O(((1+\hat\lambda)\hat\lambda)^2(2((2\hat\lambda)^{\Delta(t+1)}+1))^{2\log n}m^3(1+\log\hat\lambda))$$
$$= O(((1+\hat\lambda)\hat\lambda)^2(1+\log\hat\lambda)m^3n^{2\Delta(t+2)(1+\log\hat\lambda)+2}).$$

\subsection{Maximal independent sets and maximal $b$-matchings in the non-hierarchical framework}
\label{sec:mismaxandb}
\subsubsection{Dealing with internally disconnected classes}
As noted in Appendix~\ref{sec:adapting}, we use the non-hierarchical framework and assume bounded treewidth and degree for the chains on maximal independent sets and maximal $b$-matchings. Once we have defined these chains, we will see that partitioning the Glauber graph for each chain in the natural way will result in classes that are not necessarily internally connected. The solution will be to relax the framework conditions so that the classes need not be disjoint\textemdash but then require that every pair of overlapping classes must overlap in a large number of vertices. More precisely:
\label{sec:nonindcond}
\begin{lemma}
\label{lem:nonindnonhiercond}
Suppose a Glauber graph~$\flipg{}{G}$ satisfies the conditions of the non-hierarchical framework in Section~\ref{sec:fwcond}, except that:
\begin{enumerate}
\item The $O(1)$ classes are not necessarily disjoint.
\item Each pair of classes $\classt{}{T}$ and $\classt{}{T'}$ sharing at least one vertex shares $\Theta(1)|\classt{}{T}| = \Theta(1)|\classt{}{T'}|$ vertices.
\end{enumerate}
Then the expansion of $\flipg{}{G}$ is $\Omega(1/n^c)$, where $c = O(1)$.
\end{lemma}
\begin{proof}
The multicommodity flow construction is as in the proof of Lemma~\ref{lem:iscondexp}, except that when sending flow from $S \in \classt{}{T}$ to $S' \in \classt{}{T'} \neq \classt{}{T}$ via a path through intermediate classes, we now have some pairs of intermediate classes that share boundary vertices, instead of sharing boundary edges. The flow is the same as before, except that there is no need to send flow across a boundary in these cases.

The congestion analysis is the same as in the proof of Lemma~\ref{lem:nonindhiercond}.
\end{proof}

\subsubsection{Maximal independent sets}
\label{sec:mismaxb}

\label{sec:othersimpler}

\label{sec:mismdsexp}
We now apply the non-hierarchical framework to the chain on maximal independent sets. We first define the maximal independent set mixing problem, then show that it meets the criteria of the framework, up to the conditions in Lemma~\ref{lem:nonindnonhiercond}.

We return to the Glauber graph defined in Definition~\ref{def:misgraph}.

\begin{lemma}
\label{lem:misgraphdeg}
The maximum degree of $\flipg{MIS}{G}$ is at most $n\cdot 2^{\Delta^2+\Delta}$, where $n = |V(G)|$ and $\Delta$ is the maximum degree of any vertex in $G$.
\end{lemma}
\begin{proof}
The number of neighbors of a given maximal independent set $S$ is the same as the number of ways to choose a vertex $v$ to add to or drop from $S$, along with a subset of the vertices at distance at most two from $v$ to add or drop.
\end{proof}

\begin{definition}
\label{def:mischain}
Define the \emph{maximal independent set chain} on a graph $G$ with Glauber graph $\flipg{MIS}{G}$ as the following Markov chain (let $\Delta_M$ be the maximum degree of $\flipg{MIS}{G}$):
\begin{enumerate}
\item Let $X_0$ be an arbitrary maximal independent set in $V(\flipg{MIS}{G})$.
\item For $t \geq 0$, define $X_{t+1}$ as follows:
\subitem With probability $(1/2)(\delta(X_t)/\Delta_M)$, let $X_{t+1}$ be a neighbor in $\flipg{MIS}{G}$ of $X_t$, selected uniformly at random from the neighbors of $X_t$.
\subitem With probability $1 - (1/2)(\delta(X_t)/\Delta_M)$ let $X_{t+1} = X_t$.
\end{enumerate}
\end{definition}

For technical reasons, the following observation is necessary for obtaining a rapid mixing bound from an expansion bound on $\flipg{MIS}{G}$.
\begin{remark}
This is the standard Markov chain on $\flipg{MIS}{G}$, with appropriate self loops added in the standard way. Furthermore, by Lemma~\ref{lem:misgraphdeg}, if $G$ has bounded degree, then the degree-based weighting in Definition~\ref{def:mischain} does not cause the spectral expansion of the chain to differ by more than a polynomial factor from the edge (or vertex) expansion of $\flipg{MIS}{G}$.
\end{remark}

\subsubsection{Verification of conditions for maximal independent sets}
We show how to apply the non-hierarchical version of the framework when the treewidth and degree of $G$ are bounded. First, to satisfy Condition~\ref{cond1}, we would \emph{like} to use a partition analogous to that defined in Definition~\ref{def:ispart}: each class $\classt{MIS}{T}$ is the set of maximal independent sets that agree on their restriction to the vertex separator $X$ for $G$. However, a subtlety arises when considering the Cartesian product structure of the Glauber graphs on $A$ and $B$ within a class $T$: in the independent set Glauber graph, $\classt{IS}{T}$ was a Cartesian product of two independent set Glauber graphs $\flipg{MIS}{G_A}$ and $\flipg{MIS}{G_B}$, respectively defined on the independent sets in $A \setminus N_A(T)$ and on those in $B \setminus N_B(T)$. Here, however, the independent sets chosen in $A$ and in $B$ need to give a \emph{maximal} independent set when their union is taken with the set chosen in $X$.

It may be that the independent set in $X$ identified with $\classt{MIS}{T}$ is not maximal. For a simple example, suppose $X$ is a path of length three, consisting of vertices $u, v,$ and $w$ and edges $(u, v)$ and $(v, w)$, with $u$ having neighbors only in $A$, $w$ having neighbors only in $B$, and $v$ having only~$u$ and~$w$ as neighbors. Suppose~$\classt{MIS}{T}$ is identified with the independent set~$\{u\} \subseteq X$. Then every maximal independent set~$S \in T$ has some neighbor of~$w$ in~$B$ chosen, or else~$S$ would not be maximal. Furthermore, defining classes in this way would result in internally disconnected classes. Thus we cannot simply eliminate vertices in~$N_A(T)$ from~$A$ and~$N_B(T)$ from~$B$ and define smaller maximal independent set Glauber graphs. Instead, we define the classes\textemdash which, per the conditions of Lemma~\ref{lem:nonindnonhiercond}, need not be disjoint\textemdash as follows:

\begin{definition}
\label{def:missubclass}
Given a graph $G$ with maximal independent set Glauber graph $\flipg{MIS}{G}$ and a class $\classt{MIS}{T}$, let $U \subseteq X$ be the set of all unselected vertices in $T$ that have no neighbor selected in~$T$. For each independent subset of the vertices in $N_A(U) \cup N_B(U)$ that \emph{covers} all of $U$\textemdash that is, for each independent subset $C \subseteq N_A(U) \cup N_B(U)$ such that every~$x \in U$ has some neighbor~$z \in C$, let $\classt{MIS}{T_C}$ be the class of all independent sets in~$G$ that agree with $T$ on $X$, and that include all of the vertices in $C$.
\end{definition}

The smaller Glauber graphs on $A$ and $B$ are now independent for a given class $\classt{MIS}{T_C}$: for each~$z \in C$, remove~$z$ and all neighbors in $N_A(z) \cup N_B(z)$ from the graph, and consider the resulting maximal independent set Glauber graphs on $A \setminus (C \cup N_A(C))$ and on $B \setminus (C \cup N_B(C))$. Each class~$\classt{MIS}{T_C}$ is a Cartesian product of two such graphs. It suffices to show that this definition obeys the conditions of Lemma~\ref{lem:nonindnonhiercond}:

\begin{lemma}
\label{lem:missubflow}
Given a graph $G$ with bounded treewidth and degree and corresponding maximal independent set Glauber graph $\flipg{MIS}{G}$, the definition of classes in Definition~\ref{def:missubclass} satisfies the conditions of Lemma~\ref{lem:nonindnonhiercond}.
\end{lemma}
\begin{proof}
The Cartesian product structure of $\classt{MIS}{T_C}$ and the fact that $X$ is a balanced separator satisfy Conditions~\ref{cond5} and~\ref{cond6} of the non-hierarchical framework.

The classes do not partition $\classt{MIS}{T}$. However, there are $O(1)$ classes. The classes are also within an $O(1)$ size factor of one another. To see this, define the following mapping $f$ from the set of all maximal independent sets in $\flipg{MIS}{G}$ to the set of maximal independent sets in a class~$\classt{MIS}{T_C}$. For each maximal independent set $S \in V(\flipg{MIS}{G})$, let $S' = f(S) \in \classt{MIS}{T_C}$ be the following maximal independent set: (i) let $S'$ agree with $T$ on all vertices in $X$; (ii) let $S'$ agree with $T_C$ on all vertices in $N_A(T) \cup N_B(T)$; (iii) let $S'$ agree with $S$ on all vertices not in $X \cup N_A(T) \cup N_B(T)$ and having no neighbor in $X \cup N_A(T) \cup N_B(T)$; (iv) add vertices to $S$, if needed, to obtain maximality.

The symmetric difference $f(S) \oplus S$ is of size at most $\Delta^2|X| \leq \Delta^2(t+1) = O(1)$, where $t$ is the (bounded) treewidth of $G$ and $\Delta$ is the (bounded) degree; thus $f$ is an $O(1)$-to-1 mapping. By similar reasoning, the number of shared maximal independent sets between any two overlapping classes $\classt{MIS}{T_C}$ and $\classt{MIS}{T_{C'}}$ is at least $\Omega(1)|\classt{MIS}{T_C}|$, and the number of edges between any two adjacent classes $\classt{MIS}{T_C}$ and $\classt{MIS}{{T'}_{C'}}$ is at least $\Omega(1)|\classt{MIS}{T_C}|$. The lemma follows.
\end{proof}

\subsubsection{Maximal $b$-matchings in the non-hierarchical framework}
We return to the maximal $b$-matching Glauber graph defined in Section~\ref{sec:prelim}.

The argument that the graph is connected is similar to the proof of Lemma~\ref{lem:misgraphconn}.

It suffices to define a partition and verify the conditions. We would \emph{like} to identify each class~$\classt{MBM}{T}$ with the chosen subset of the edges that have at least one endpoint in the small balanced separator~$X$. However, as with maximal independent sets, the maximality requirement introduces non-independent subproblems. To modify the definition of the classes, we first need to introduce the notion of a \emph{saturated} vertex:

\begin{definition}
\label{def:vsat}
Given a $b$-matching in a graph $G$, consider a vertex $v$ \emph{saturated} if $b(v)$ edges incident to $v$ are selected in the matching.
\end{definition}

It may be that a vertex $v \in X$ is not saturated in a maximal $b$-matching, and thus the choice of edges inducing~$\classt{MBM}{T}$ does not saturate~$v$. In this case, we have a constraint on the subproblems in~$A$ and~$B$. Namely, it must be that some neighbor of~$v$, $u \in N(X)$, is saturated, or else the edge~$(u, v)$ could be added to the matching. We use this fact to define the subclasses of a class~$\classt{MBM}{T}$:

\begin{definition}
\label{def:bsubclass}
Given a graph~$G$ with separator~$X$, maximal $b$-matching Glauber graph~$\flipg{MBM}{G}$, and a set~$T$ of edges selected whose endpoints all lie in~$X$, let~$U \subseteq X$ be the set of unsaturated vertices in~$X$ induced by~$T$; let~$C$ be a minimal set of edges such that, after adding~$C$ to~$T$, some neighbor of~$v$ is saturated for every~$v \in U$. Define the class~$\classt{MBM}{T_C}$ as the set of all maximal $b$-matchings in~$T$ that contain all edges in~$C$.
\end{definition}

As in Section~\ref{sec:mismdsexp}, these classes are not equivalence classes, because they overlap. Again, however, each $b$-matching belongs to at most~$O(1)$ subclasses, and thus this overlap does not interfere with the proof.

We now verify that this definition of classes satisfies Lemma~\ref{lem:nonindnonhiercond}:
\begin{lemma}
\label{lem:bsubflow}
Given a graph $G$ with bounded treewidth and degree, corresponding maximal $b$-matching Glauber graph $\flipg{MBM}{G}$, the division into classes as in Definition~\ref{def:bsubclass} satisfies the conditions of Lemma~\ref{lem:nonindnonhiercond}.
\end{lemma}
\begin{proof}
The argument is similar to the proof of Lemma~\ref{lem:missubflow}: again we have a Cartesian product structure in each class~$\classt{MBM}{T_C}$\textemdash where the resulting maximal $b$-matching subproblems on~$A$ and~$B$ result from (i) removing each edge $(u, w) \in C$ from~$G[A]$ and~$G[B]$, and (ii) decreasing~$b(u)$ and~$b(w)$ accordingly.

The number of classes is~$O(1)$, due to the bounded treewidth and bounded degree of~$G$. The classes differ by an~$O(1)$ size factor, and the overlaps are large; the argument, along with the resulting flow, is similar to that in the proof of Lemma~\ref{lem:missubflow}.
\end{proof}

\subsubsection{Specific mixing upper bounds for maximal independent sets and maximal $b$-matchings}
For the derivation of the specific bounds stated in Theorem~\ref{thm:mindsetmix}, we apply Lemma~\ref{lem:nonhierexact}, with the modification that the $\mathcal{E}_{\min}$ term must be replaced by the term $\min\{\mathcal{E}_{\min}, \mathcal{O}_{\min}\}$, where $\mathcal{O}_{\min}$ is the size of the smallest overlap between a pair of classes that share at least one vertex. For maximal independent sets, $\mathcal{E}_{\min} \geq N/2^{7\Delta^6(t+1)}$ and $\mathcal{O}_{\min} \geq N/2^{3\Delta^2(t+1)}$, so $\min\{\mathcal{E}_{\min}, \mathcal{O}_{\min}\} \geq N/2^{7\Delta^6(t+1)}$. We also gain at each level of induction an additional factor of $K \leq 2^{(\Delta+1)(t+1)}$ due to overlaps. Combining this with the fact that $\Delta_{\mathcal{M}} \leq 2^{3\Delta^2}n$ and $N \leq 2^n$ gives a total mixing bound of

$$O((2\cdot 2^{7\Delta^6(t+1)}\cdot 2^{(\Delta+1)(t+1)}+1)^{2\log n}\cdot 2^{6\Delta^2}n^3)$$
$$= O(2^{6\Delta^2}n^{2(t+1)(7\Delta^6+\Delta+1)+7}),$$
as claimed. (The $\log(3/2)$ term in the theorem statement comes from the fact that the base of the log in the induction is 3/2.)

A similar argument for maximal $b$-matchings gives the result claimed in Theorem~\ref{thm:mindsetmix}, with

$\Delta_{\mathcal{M}} \leq 2^{6\Delta^2}m$, $N \leq 2^m$, $K \leq 2^{3\Delta^2(t+1)},$ $\mathcal{E}_{\min} \geq N/2^{8\Delta^7(t+1)}$, and $\mathcal{O}_{\min} \geq N/2^{4\Delta^3(t+1)}$.

I.e., we have
$$O((2\cdot 2^{8\Delta^7(t+1)}\cdot 2^{3\Delta^2(t+1)}+1)^{2\log n}\cdot 2^{12\Delta^2}m^3)$$
$$= O(2^{12\Delta^2}m^3n^{2(t+1)(8\Delta^7+3\Delta^2)+4}).$$

\section{Deferred Proof Details}
\label{sec:appdetails}
\lemiscondfacts*
\begin{proof}
For claim 2, consider the class~$\classt{IS}{T_r}$ whose vertex set in~$X$ is the empty set, and consider any class $\classt{IS}{T} \neq \classt{IS}{T_r}$. $\classt{IS}{T_r}$ consists of the set of all pairs~$(S_A, S_B)$, where~$S_A$ is an independent set in~$A$, and~$S_B$ is an independent set in~$S_B$. $\classt{IS}{T}$ consists of the set of all pairs~$(S_A', S_B')$, where~$S_A'$ is an independent set in~$A \setminus N_A(T)$, and~$S_B'$ is an independent set in~$B \setminus N_B(T)$.

Every independent set~$S_A'$ in~$A \setminus N_A(T)$ is also an independent set in~$A$ (and the situation is the same for~$S_B'$), so a trivial injective mapping exists from the sets in~$\classt{IS}{T}$ to the sets in~$\classt{IS}{T_r}$. For the other direction, consider the mapping $f: \mathcal{P}(A) \rightarrow \mathcal{P}(A \setminus N_A(T))$ that sends every independent set~ $S_A \subseteq A$ to its restriction $S_A' = S_A \setminus N_A(T)$. Because the degree~$\Delta$ of~$G$ is bounded, $|N_A(T)| \leq t\Delta = O(1),$ and thus~$f$ is at worst a $2^{t\Delta} = O(1)$-to-one mapping. This shows that the classes differ in size by a factor of~$O(1)$, proving the lemma.

For claim 3, each edge $(S, S')$ between independent sets $S \in \classt{IS}{T}$ and $S' \in \classt{IS}{T'} \neq \classt{IS}{T}$ consists of adding or removing a single vertex $v \in X$. $S$ has no other moves to independent sets in $\classt{IS}{T'}$.

For claim 4, $T$ and $T'$ differ by exactly one vertex; call it $v$. (Or else no move could exist between $\classt{IS}{T}$ and $\classt{IS}{T'}$.) Suppose $v \in T$ and $v \notin T'$; then every independent set in $\classt{IS}{T}$ has a move to some independent set in $\classt{IS}{T'}$. (See Figure~\ref{fig:isflip}.) Thus the number of edges from $\classt{IS}{T}$ to $\classt{IS}{T'}$ is~$|\classt{IS}{T}|$; the lemma now follows from Lemma~\ref{lem:iscondclasssizes}.
\end{proof}

\lemmisgraphconn*
\begin{proof}
Let $S \neq S'$ be maximal independent sets, and consider the symmetric difference $S \oplus S'$: if $|S \oplus S'| > 0$, choose some $v \in S' \setminus S$. Obtain a new set $S''$ by adding $v$ to $S$ and removing all neighbors of $v$ from $S$, then greedily adding neighbors of neighbors of $v$ until an maximal independent set is obtained. Repeat this process with a new vertex $v' \in S' \setminus S''$, and so on, for every vertex in $S' \setminus S$, obtaining a sequence of sets $S_1 = S, S_2 = S'', S_3, \dots, S_k$. Crucially, once a vertex $v$ is selected from $S'$ in this process, giving set $S_i$, we have $v \in S_j$ for all $i \leq j \leq k$. This is because the only way for a vertex to be removed in the process is for one of its neighbors to be selected from $S'$. However, since $S'$ is an independent set, no neighbor of $v$ belongs to $S'$.

Thus we have $S_k = S'$, proving that there is a path in $\flipg{MIS}{G}$ between every pair of maximal independent sets.
\end{proof}

\bibliography{references}

\end{document}